\documentclass[sigplan,10pt]{acmart}
\renewcommand\footnotetextcopyrightpermission[1]{} 
\usepackage{balance}

\usepackage{ifthen}
\usepackage{amssymb}
\usepackage{tikz}
\usepackage{verbatim}
\usepackage{listings}
\usepackage{hyperref}
\usepackage{stmaryrd}
\usetikzlibrary{shapes,arrows,shadows}

\newcommand{\seppred}[2]{\ensuremath{#1{\langle}#2{\rangle}}}

\newcommand{\report}[1]{ }

\newcommand{\acm}[1]{ }

\newcommand{\hide}[1]{}
\newcommand{\hideie}[1]{}
\newcommand{\simp}{{\bf ;}}

\newcommand{\heap}{\ensuremath{\kappa}}

\newcommand{\rulen}[1]{\ensuremath{{\bf \scriptstyle [#1]}}}

\newcommand{\pred}[2]{\seppred{\btt{#1}}{#2}}

\usepackage{float}
\floatstyle{boxed}
\restylefloat{figure}

\def\int{\code{int}}

\def\true{\code{true}\,}
\def\false{\code{false}\,}

\newcommand{\code}[1]{{\small {\ensuremath{\tt #1}}}}

\newcommand{\btt}[1]{{\ensuremath{\tt #1}}}

\newcommand{\wnnay}[1]{}

\newcommand{\locnay}[1]{}

\newcommand{\tmnay}[1]{}

\newcommand{\cdnay}[1]{}
\newcommand{\nodo}[1]{}

\newtheorem{prop}{Proposition}

\def\S{\Sigma}

\definecolor{light-gray}{gray}{0.85}

\newcommand{\infrule}[1]{\underline{\rulen{#1}}}

\usepackage{graphicx}
\usepackage{array}
\usepackage{url}
\usepackage{xcolor}

\usepackage{float}
\usepackage{wrapfig}
\usepackage[linesnumbered,ruled,boxed]{algorithm2e}

\usepackage{time}
\usepackage{datetime}
\usepackage{amsmath}
\usepackage{amssymb}
\usepackage{lscape}
\usepackage{time}
\usepackage{datetime}
\usepackage{proof}

\usepackage{framed}

\usepackage{enumitem}

\usepackage{color}
\makeatletter 
\let\c@lofdepth\relax 
\let\c@lotdepth\relax 
\makeatother 
\usepackage[raggedright,small,sf,SF,hang]{subfigure}

\usepackage{float}
\usepackage{wrapfig}

\usepackage{extarrows}  

\usepackage{stmaryrd}   

\begin{document}



\renewcommand{\baselinestretch}{0.92} 
\newcommand{\clpr}{CLP(${\cal R}$)}
\newcommand{\np}{${\cal NP}$~}
\newcommand{\TR}{${\cal TR}$}
\newcommand{\CTR}{${\cal CTR}$}
\newcommand{\SR}{${\cal SR}$~}
\newcommand{\del}{{\em delta}}
\newcommand{\nil}{~{\rm nil}~}
\newcommand{\I}{{\cal I}~}

\newcommand{\ignore}[1]{}

\newcommand{\stuff}[1]{
        \begin{minipage}{6in}
        {\tt \samepage
        \begin{tabbing}
        \hspace{3mm} \= \hspace{3mm} \= \hspace{3mm} \= \hspace{3mm} \= \hspace{3mm} \= \hspace{3mm} \=
\hspace{3mm} \= \hspace{3mm} \= \hspace{3mm} \= \hspace{3mm} \= \hspace{3mm} \= \hspace{3mm} \= \hspace{3mm} \= \kill
        #1
        \end{tabbing}
       }
        \end{minipage}
}

\newcommand{\mystuff}[1]{
        \begin{minipage}[b]{6in}
        {\tt \samepage
        \begin{tabbing}
        \hspace{3mm} \= \hspace{3mm} \= \hspace{3mm} \= \hspace{3mm} \= \hspace{3mm} \= \hspace{3mm} \=
\hspace{3mm} \= \hspace{3mm} \= \hspace{3mm} \= \hspace{3mm} \= \hspace{3mm} \= \hspace{3mm} \= \hspace{3mm} \= \kill
        #1
        \end{tabbing}
       }
        \end{minipage}
}

\newcommand{\newmystuff}[1]{
        \begin{minipage}[t]{6in}
        {\tt \samepage
        \begin{tabbing}
        \hspace{3mm} \= \hspace{3mm} \= \hspace{3mm} \= \hspace{3mm} \= \hspace{3mm} \= \hspace{3mm} \=
\hspace{3mm} \= \hspace{3mm} \= \hspace{3mm} \= \hspace{3mm} \= \hspace{3mm} \= \hspace{3mm} \= \hspace{3mm} \= \kill
        #1
        \end{tabbing}
       }
        \end{minipage}
}

\newcommand{\Rule}[2]{\genfrac{}{}{0.5pt}{}
{{\setlength{\fboxrule}{0pt}\setlength{\fboxsep}{3mm}\fbox{$#1$}}}
{{\setlength{\fboxrule}{0pt}\setlength{\fboxsep}{3mm}\fbox{$#2$}}}}

\newcommand{\rct}{\mbox{\it clp}}
\newcommand{\lif}{\ {\tt:\!\!-} \ }
\newcommand{\ctbl}{\,{\mbox{\footnotesize$|$}\tt-\!\!\!\!>}\,}
\newlength{\colwidth}
\setlength{\colwidth}{.47\textwidth}


\newcommand{\eqdef}{\stackrel{\rm def}{=}}
\newcommand{\sys}{\mbox{\bf sys}}
\newcommand{\var}{\mbox{\frenchspacing \it var}}
\newcommand{\lfp}{\mbox{\it \frenchspacing lfp}}
\newcommand{\gfp}{\mbox{\it \frenchspacing gfp}}
\newcommand{\assert}{\mbox{\it \frenchspacing assert}}
\newcommand{\buffer}{\mbox{\it \frenchspacing Buffer}}
\newcommand{\clpif}{\mbox{\tt :-}}


\newtheorem{myproof}{Proof Sketch}

\newcommand{\QED}{\nolinebreak\hskip 1em
        \framebox[0.5em]{\rule{0ex}{1.0ex}}}

\newcommand{\close}{{\frenchspacing close }}
\newcommand{\rules}{{\frenchspacing rules }}
\newcommand{\lhs}{{\frenchspacing lhs }}
\newcommand{\rhs}{{\frenchspacing rhs }}
\newcommand{\wrt}{{\frenchspacing wrt. }}
\newcommand{\fold}{\mbox{\it\frenchspacing old}}
\newcommand{\exactunfold}{\mbox{\it\frenchspacing exactunfold}}

\newcommand{\mgu}{{\frenchspacing mgu }}
\newcommand{\size}{{\frenchspacing size }}
\newcommand{\obs}{{\frenchspacing obs }}
\newcommand{\trace}{{\frenchspacing trace }}

\newcommand{\xxx}{\mbox{\Large $\vartriangleright$}}

\newcommand{\undeniable}{\mbox{\Large $\vartriangleright$}\hspace{-8pt}\raisebox{2pt}{\tiny u}~~}
\newcommand{\inevitable}{\mbox{\Large $\vartriangleright$}\hspace{-8pt}\raisebox{2pt}{\tiny i}~~\,}

\newcommand{\reachable}{\mbox{\Large $\vartriangleright$}\hspace{-9pt}\raisebox{1pt}{\tiny *}~~\,}

\newcommand{\lbr}{\mbox{$[\![$}}
\newcommand{\rbr}{\mbox{$]\!]$}}
\newcommand{\ct}[1]{\mbox{\lbr$\vec{#1}$\rbr}}
\newcommand{\cs}[1]{\mbox{\lbr#1\rbr}}
\newcommand{\clp}[1]{{\it \frenchspacing clp}(\mbox{$#1$})}
\newcommand{\pp}[1]{\mbox{{\color{blue}{$\langle$#1$\rangle$}}}}

\newcommand{\ao}{\mbox{${\cal A}$}}
\newcommand{\co}{\mbox{${\cal C}$}}
\newcommand{\po}{\mbox{${\cal P}$}}

\newcommand{\tA}{{\tilde{A}}}
\newcommand{\tB}{{\tilde{B}}}
\newcommand{\tC}{{\tilde{C}}}

\newcommand{\cts}{\mbox{$\mapsto$}}
\newcommand{\sepimp}{\mbox{$-\!*$}}

\newcommand{\arr}[3]{\mbox{$\langle \mbox{#1,#2,#3} \rangle$}}
\newcommand{\triple}[3]{\langle{#1},{#2},{#3}\rangle}
\newcommand{\oldtriple}[3]{\lbrace #1 \rbrace ~#2~ \lbrace #3 \rbrace}
\newcommand{\aquadruple}[4]{\langle{#1},{#2},{#3},{#4}\rangle}
\newcommand{\baeq}[2]{\mbox{$=_{\[#1 .. #2\]}$}}

\newcommand{\ite}[3]{\mbox{\textit{ite}}({#1},{#2},{#3})}
\newcommand{\ptab}{~~~~}

\newlength{\vitelen}
\settowidth{\vitelen}{\mbox{\textit{ite}}(}
\newcommand{\vite}[3]{\begin{array}[t]{l}\mbox{\textit{ite}}({#1},\\
\hspace{\vitelen}{#2},\\
\hspace{\vitelen}{#3}
    \end{array}}

\newcommand{\Hfx}{H_{\!f}}
\newcommand{\Hx}{H_{\!1}}
\newcommand{\Hxx}{H_{\!2}}
\newcommand{\Pf}{P_{\!f}}
\newcommand{\Pz}{P_{\!0}}
\newcommand{\Px}{P_{\!1}}

\newcommand{\Ix}{I_{\!1}}
\newcommand{\Ixx}{I_{\!2}}
\newcommand{\Jf}{J_{\!f}}
\newcommand{\Jx}{J_{\!1}}
\newcommand{\Jxx}{J_{\!2}}

\newcommand{\witness}{\mbox{$\omega$}}
\newcommand{\transformer}{\mbox{$\Delta$}}

\newcommand{\func}[1]{\mbox{\textsf{#1}}}

\floatstyle{boxed}
\restylefloat{figure}

%

\renewcommand\labelitemi{$\bullet$}
\renewcommand\labelitemii{\normalfont\bfseries --}

\renewcommand\floatpagefraction{.9}
\renewcommand\topfraction{.9}
\renewcommand\bottomfraction{.9}
\renewcommand\textfraction{.1}   
\setcounter{totalnumber}{50}
\setcounter{topnumber}{50}
\setcounter{bottomnumber}{50}

\raggedbottom

\makeatletter
\newcommand{\manuallabel}[2]{\def\@currentlabel{#2}\label{#1}}
\makeatother

\newcounter{chapcount}
\newcommand{\chapcountreset}{\setcounter{chapcount}{0}}
\chapcountreset

\newcounter{excount}
\newcommand{\exreset}{\setcounter{excount}{0}}
\exreset
\newcommand{\newexample}[1]{\addtocounter{excount}{1}
{\vspace{2mm} \noindent \mbox{{\scriptsize EXAMPLE} \examplecount~\emph{#1}:}}}

\newcommand{\examplecount}{\arabic{excount}}

\newcommand{\exlabel}[1]{\manuallabel{#1}{\examplecount}}

\newcommand{\todo}[1]{{{\color{red} {\bf TODO:} \textit{#1}}}}

\newcommand{\hcomment}[1]{{\color{magenta} [#1]}}

\newcommand{\memoed}{\mbox{\func{memoed}}}
\newcommand{\memoize}{\mbox{\func{memoize}}}

\newcommand{\memotable}{\mbox{\pred{Table}}}
\newcommand{\wpc}{\mbox{$\func{pre}$}}

\newcommand{\interp}{\mbox{$\pred{Intp}$}}
\newcommand{\emanate}{\mbox{\func{outgoing}}}
\newcommand{\absiteration}{\mbox{\func{loop\_end}}}
\newcommand{\step}{\mbox{\func{TransStep}}}
\newcommand{\compress}{\mbox{\func{JoinVertical}}}
\newcommand{\join}{\mbox{\func{JoinHorizontal}}}

\newcommand{\program}{\mbox{$\cal P$}}
\newcommand{\N}{\mbox{$\cal N$}}

\newcommand{\Assign}{\mbox{:=}}
\newcommand{\pair}[2]{\langle{#1},{#2}\rangle}
\newcommand{\tuple}[3]{\langle{#1},{#2},{#3}\rangle}
\newcommand{\quadruple}[4]{[{#1},{#2},{#3},{#4}]}
\newcommand{\fivetuple}[5]{[{#1},{#2},{#3},{#4},{#5}]}

\newcommand{\ckeyword}[1]{{\color{red} \textbf{#1}}}

\newcounter{pppcount}
\newcommand{\pppreset}{\setcounter{pppcount}{0}}
\newcommand{\ppp}{\refstepcounter{pppcount}\mbox{{\color{blue}$\langle\arabic{pppcount}\rangle$}}}

\newcommand{\pppred}{\refstepcounter{pppcount}\mbox{{\color{red}$\texttt{\arabic{pppcount}}$}}}

\newcommand{\mnote}[1]{\marginpar{\color{red}{#1}}}

\newcommand{\hiepalert}[1]{{\small {\color{red}{#1}}}}
\newcommand{\hiepassert}[1]{{\small {\color{blue}{#1}}}}

\newcommand{\resource}{\mbox{\textsf r}}
\newcommand{\timing}{\mbox{\textsf t}}

\newcommand{\summarize}{\mbox{\cal S}}


\newcommand{\myiff}{\mbox{\texttt{iff}}}

\newcommand{\assume}[1]{\mbox{\textsf{assume(#1)}}}
\newcommand{\assign}[2]{\mbox{\textsf{#1 := #2}}}

\newcommand{\exec}{\mbox{\textsf{exec}}}

\newcommand{\transition}[3]{#1~\xlongrightarrow[]{#3}~#2}
\newcommand{\shorttransition}[2]{#1~\xrightarrow[]{}~#2}
\newcommand{\shorttrans}{\rightarrow}
\newcommand{\translabel}[1]{\xlongrightarrow[]{#1}}
\newcommand{\loc}{\mbox{$\ell$}}
\newcommand{\locations}{\mbox{${\cal L}$}}

\newcommand{\transsystem}{\mbox{$\mathcal{P}$}}
\newcommand{\newtranssystem}{\mbox{$\mathcal{G}$}}

\newcommand{\symstate}{\mbox{$s$}}
\newcommand{\pci}[1]{\mbox{$\loc_{#1}$}}
\newcommand{\pc}{\mbox{\loc}}
\newcommand{\pcend}{\mbox{$\loc_{\textsf{end}}$}}
\newcommand{\pcerror}{\mbox{$\loc_{\textsf{error}}$}}
\newcommand{\pcstart}{\mbox{$\loc_{\textsf{start}}$}}
\newcommand{\pathcond}{\mbox{$\Pi$}}
\newcommand{\pathcondbar}{\mbox{$\overline{\Pi}$}}
\newcommand{\storebar}{\mbox{$\overline{h}$}}
\newcommand{\symstatebar}{\mbox{$\overline{\symstate}$}}
\newcommand{\mapstatetoformula}[1]{\mbox{$\llbracket {#1} \rrbracket$}}

\renewcommand{\path}{\mbox{$\theta$}}

\newcommand{\typevar}{\mbox{\emph{Vars}}}
\newcommand{\typesymvar}{\mbox{\emph{SymVars}}}
\newcommand{\typeop}{\mbox{\emph{Ops}}}
\newcommand{\typefo}{\mbox{\emph{FO}}}
\newcommand{\typeterms}{\mbox{\emph{Terms}}}
\newcommand{\typestate}{\mbox{\emph{States}}}
\newcommand{\typesymbstate}{\mbox{\emph{SymStates}}}
\newcommand{\typesympath}{\mbox{\emph{SymPaths}}}
\newcommand{\typebool}{\mbox{\emph{Bool}}}
\newcommand{\typeint}{\mbox{\emph{Int}}}
\newcommand{\typenat}{\mbox{\emph{Nat}}}
\newcommand{\typekeys}{\mbox{$\mathcal{K}$}}
\newcommand{\typevoid}{\mbox{\emph{Void}}}

\newcommand{\eval}[2]{\llbracket {#1} \rrbracket_{#2}}
\newcommand{\define}{\mbox{~$\triangleq$~}}
\newcommand{\unknown}{\mbox{\textsf{$\cdot$}}}

\newcommand{\Intpsymbol}{\mbox{$\overline{\Psi}$}}
\newcommand{\InvariantFunc}{\mbox{\textsf{invariant}}}
\newcommand{\InvariantSym}{\mbox{$\mathcal{I}$}}
\newcommand{\ConflictSym}{\mbox{$\mathcal{C}$}}
\newcommand{\ContextSym}{\mbox{$\mathcal{O}$}}
\newcommand{\modifies}{\mbox{\textsf{\textsc{Modifies}}}}
\newcommand{\havoc}{\mbox{\textsf{\textsc{Havoc}}}}
\newcommand{\getvars}{\mbox{\textsf{var}}}

\newcommand{\Enclose}{\mbox{\textsc{enclose}}}
\newcommand{\SEnclose}{\mbox{\texttt{enclose}}}
\newcommand{\Evolve}{\mbox{\textsc{evolve}}}
\newcommand{\SEvolve}{\mbox{\texttt{evolve}}}




\floatsep 3mm plus 1mm minus 1mm
\dblfloatsep 6pt plus 2pt minus 2pt 
\textfloatsep 3mm plus 1mm minus 1mm
\dbltextfloatsep 6pt plus 2pt minus 4pt
%


\title{Automatic Reasoning on Recursive Data Structures with Sharing}


\author{Duc-Hiep Chu}
\affiliation{IST Austria}
\email{duc-hiep.chu@ist.ac.at}   

\author{Joxan Jaffar}
\affiliation{National University of Singapore}          
\email{joxan@comp.nus.edu.sg}      





\newcommand{\jj}[1]{\fbox{{\color{magenta} \textbf{#1}}}}

\newcommand{\HL}[0]{\mathcal{HL}}
\newcommand{\HH}[0]{\mathcal{HL}}

\newcommand{\vars}[0]{\mathit{vars}}
\newcommand{\fp}[0]{\mathit{F\!P}}
\newcommand{\wtriple}[3]{\{#1\}~#2~\{#3\}_W}
\newcommand{\vtriple}[3]{\begin{array}{c}\{#1\}\\#2\\\{#3\}\end{array}}
\newcommand{\dom}[0]{\mathit{dom}}
\newcommand{\set}[1]{\mathsf{#1}}
\renewcommand{\code}[1]{\llbracket #1 \rrbracket}

\renewcommand{\prop}[0]{\Longrightarrow}
\renewcommand{\simp}[0]{\Longleftrightarrow}

\newcommand{\hiepand}{\texttt{,~}}

\newcommand{\Null}[0]{{\textsf{null}}}
\newcommand{\lef}[0]{{le\!f\!t}}
\newcommand{\underscore}[0]{{\underline{\LARGE\ }}}
\newcommand{\footprint}[0]{{\textsf{footprint}}}

\newcommand{\vv}[0]{\mathcal{V}}
\renewcommand{\heap}[0]{\mathcal{H}}
\newcommand{\heapg}[0]{\mathcal{M}}
\newcommand{\ghost}[0]{\mathcal{G}}

\newcommand{\hemp}[0]{\mathsf{emp}}
\newcommand{\hone}[2]{(#1 \mapsto #2)}
\newcommand{\hsep}[0]{{\symbol{42}}}
\newcommand{\heq}[0]{\bumpeq}
\newcommand{\hin}[3]{\mathsf{in}(#1, #2, #3)}
\newcommand{\hreach}[0]{\mathsf{reach}}
\newcommand{\hassign}[0]{\mathsf{assign}}
\newcommand{\hdom}[0]{\mathsf{dom}}
\newcommand{\hcontain}[0]{\rhd}

\newcommand{\update}[0]{\mathsf{update}}
\newcommand{\enclose}[0]{\mathsf{en\_upd}}
\newcommand{\fram}[0]{\mathsf{frame}}
\newcommand{\evolve}[0]{\mathsf{evolve}}

\newcommand{\cinit}[0]{\phi_\mathit{init}}
\newcommand{\cinv}[0]{\phi_\mathit{inv}}
\newcommand{\cexit}[0]{\phi_\mathit{exit}}
\newcommand{\astate}[0]{\sigma}
\newcommand{\ainit}[0]{\mathsf{precond}}
\newcommand{\ainv}[0]{\mathsf{invariant}}
\newcommand{\aexit}[0]{\mathsf{postcond}}

\newcommand{\nodeskel}[0]{\mathsf{node}_\textsc{skel}}
\newcommand{\closedskel}[0]{\mathsf{closed}_\textsc{skel}}
\newcommand{\nodenode}[0]{\mathsf{node}_\textit{node}}
\newcommand{\closednode}[0]{\mathsf{closed}_\textit{node}}

\newcommand{\ditto}[0]{$''$}

\newcommand{\cmark}[0]{\ding{51}}
\newcommand{\xmark}[0]{\ding{55}}

\newcommand{\RN}[1]{\uppercase\expandafter{\romannumeral #1\relax}}

\newcommand{\one}[2]{(#1 \mapsto #2)}

\newcommand{\nextpointer}{{\sf {nxt}}}
\newcommand{\lseg}{{\sf {ls}}}
\newcommand{\listn}{{\sf {list}}}
\newcommand{\lslast}{{\sf {list\_last}}}
\newcommand{\lsapp}{{\sf {list\_append}}}
\newcommand{\lsegleft}{${\sf \widehat{ls}}$}
\newcommand{\lsegleftmath}{{\sf \widehat{ls}}}
\newcommand{\sortedls}{{\sf {sorted\_ls}}}
\newcommand{\sortedlist}{{\sf {sorted\_list}}}

\makeatletter
\DeclareRobustCommand*\cal{\@fontswitch\relax\mathcal}
\makeatother

\begin{abstract}

We consider the problem of automatically verifying programs which
manipulate arbitrary data structures.  Our specification language is
expressive, contains a notion of \emph{separation}, and thus enables a
precise specification of \emph{frames}.  The main contribution then is
a program verification method which combines strongest postcondition
reasoning in the form symbolic execution, unfolding recursive
definitions of the data structure in question, and a new frame rule to
achieve \emph{local reasoning} so that proofs can be compositional.
Finally, we present an implementation of our verifier, and demonstrate
automation on a number of representative programs.  In particular, we
present the first automatic proof of a classic graph marking
algorithm, paving the way for dealing with a class of programs which
traverse a complex data structure.

\end{abstract}


\maketitle

\section{Introduction}
\label{sec:intro}
Formal reasoning about programs which manipulate dynamically allocated
data structures is challenging; it is more so when data structures
that (1) have ``unrestricted'' sharing, e.g., as in an arbitrary
graph; and (2) are \emph{recursive}, that is, their formal definition
is represented by a recursively defined relation.  A key technical
difficulty is that ``deep aliasing'' prevents us from reasoning in
isolation over parts of these data structures, a.k.a. \emph{local
reasoning}, because changes to one part of the structure (say, the
left child of a graph) can affect other parts (the right child or its
descendants) that may point into it.
Consequently, there is no established systematic method 
than can \emph{automatically} verify programs
which manipulate such data structures, even for routine
textbook graph algorithms.    


In traditional Hoare logic, there is a \emph{frame rule} (CFR), which
allows an assertion that does not mention heap variables or
pointers, to be ``framed'' through a program fragment.  Augmenting
this rule to accommodate 
recursively defined predicates has be used from as early as
$1982$ \cite{morris82tfpm,bornat00mpc}.

\begin{proposition}[Classic Frame Rule] 
\ \\ \vspace*{-4mm}
\begin{center}
$\begin{array}{c}
\infer{\oldtriple{\phi \wedge \pi}{P}{\psi \wedge \pi}}
{\oldtriple{\phi}{P}{\psi}}
\end{array}Mod(P) \cap FV(\pi) = \emptyset$ \hfill {\sc (CFR)}
\end{center}

\noindent
where $Mod(P)$ denotes the variables that $P$ modifies,
and $FV(\pi)$ denotes the free variables of $\pi$. $~~~\Box$
\end{proposition}

\floatstyle{plain}
\restylefloat{figure}

\begin{wrapfigure}{r}{0.20\textwidth}
\vspace{-6mm}
$\infer{\oldtriple{\phi ~\hsep~ \pi}{P}
{\psi ~\hsep~ \pi}}{\oldtriple{\phi}{P}{\psi}}$\hfill {\sc (SFR)}
\vspace{-6mm}
\end{wrapfigure}

It was Separation Logic \cite{ohearn01local,reynolds02separation} (SL)
which made a significant advance in verifying programs that manipulate
recursive data structures such as linked-lists or trees.  Two key
ideas here are: associating a predicate with a notion of \emph{heap},
and composing predicates with the notion of \emph{separating
conjunction} of heaps.  As a result, SL has an extremely elegant frame
rule (SFR): when a program fragment $P$ is ``enclosed'' in some heap,
then any formula $\pi$ whose ``footprint'' is separate from this heap
can be ``framed'' through the fragment.  This notion of separation is
indicated by the ``separating conjunction'' operator ``$*$'' which
states that the footprints of its two operands (which are logical
predicates) are disjoint.  Note that the validity of the triple
$\oldtriple{\phi}{P}{\psi}$ entails that all heap accesses in $P$,
read or write, are confined to the implicit heap of $\phi$, or to
fresh addresses.  This provides for truly local reasoning, because the
proof of $P$ is done \emph{without any prior knowledge} about the
predicate $\pi$.

There have been significant advancements in automating SL and its
variants to deal with recursive data-structures (without sharing)
such as linked-lists or trees 
\cite{nguyen10shape3,wies13cav,qiu13dryad,piskac14cav,pek14pldi,chu15pldi}.
However, we cannot directly use SL frame rule to deal with programs
that manipulate data structures with sharing because such structures
cannot easily be massaged into the form $\phi ~\hsep~ \pi$: for
example, the left and right descendants of a graph node often are not
disjoint.

The recent work \cite{hobor13ramify} was an important contribution toward
the sharing problem in SL.   It introduced a new
``ramification'' rule which allowed for shared structures.
However, the preconditions to using this rule, including the
use of ``magic wand'' operators, are complex.
Consequently, though we now have a systematic approach to formally
reason about shared structures,
we are not yet closer to an \emph{automated} method.
We elaborate on \cite{hobor13ramify} in Section \ref{sec:related}.

After SL, we have the method of \emph{dynamic frames}
\cite{kassios06dynamic} (DF), and later, the refinement to
\emph{implicit} dynamic frames (IDF) ~\cite{smans09implicit}.
Essentially, a DF is a mathematical expression describing a (super)set
of memory locations that a method refers to, a.k.a. the ``footprint''
of the method.  It is used in the specification phase of verification:
together with the provided pre and postcondition of a method, the DF
represents the ``modifies'' property of the method. 
The big advantage of DF/IDF over SL is expressiveness, where frames can be
specified at a higher resolution.
However, while DF/IDF serve to automate local reasoning, they do not
directly address \emph{recursive data structures with sharing}.  A key
reason for this is that recursion typically embeds \emph{reachability}
properties, and such properties are not typically accommodated by SMT
solvers that are often used in DF/IDF systems.  As circumstantial
evidence for this, note that a challenge problem, marking a graph, was
first systematically proved, albeit manually, in \cite{hobor13ramify}.
(We will show later a first systematic automatic proof.)  We elaborate
on DF/IDF in Section \ref{sec:related}.


In this paper, we propose a framework that enables \emph{fully
automatic} local reasoning on recursive data structures with sharing.
The key features of our framework that account for the new level of
automation are as follows:

\vspace{1mm}
\noindent
{\bf Strongest Postcondition Transform:}
Our assertion language, from \cite{duck13heaps},
contains explicit subheaps definable by predicates,
and separation is expressed by subheap disjointness.
Thus we can use traditional conjunction of formulas, 
as opposed to separating conjunction.
\ignore{
which subheaps may be explicitly defined within
predicates \cite{duck13heaps}, and the effect of separation obtained
by specifying that certain heaps are disjoint. As a result, the
``overloaded meaning'' of the separation conjunction is removed and
predicates are conjoined in the traditional way.  } Again
from \cite{duck13heaps}, we now have a \emph{strongest postcondition}
transform (adapted and presented in Section \ref{sec:symexec}) of heap
formulas.  What is new in this paper is to couple the transform with
 the reasoning of recursive predicates, and importantly, frame reasoning.

\vspace{1mm}
\noindent
{\bf Framing of Recursive Predicates with CFR:} We use a distinguished
heap variable $\heapg$ to represent the global heap memory, and extend
\cite{duck13heaps} by forcing all subheaps
appearing in a recursive predicate and its definition to
be \emph{ghost} variables.  A subheap $\heap$ is explicitly
constrained to be part of $\heapg$ via a ``heap reality'' constraint
$\heap \sqsubseteq \heapg$.  (E.g., an assertion stating a linked-list
rooted at $x$ will be ${\tt list}(\heap,
x) \wedge \heap \sqsubseteq \heapg$, instead of just ${\tt list}(x)$
as in SL.)  An important consequence is that, except for heap reality
constraints, other constraints, e.g. ${\tt list}(\heap, x)$, are
applicable to the classical frame rule (CFR).

\vspace{1mm}
\noindent
{\bf Framing of Heap Reality Constraints:} 
Our main contribution is a new frame rule to address framing of heap
reality constraints (Section \ref{sec:framerule}).  Intuitively, if
all the updates of the program fragment $P$ are confined within the
heap $\heap_1$ (or to fresh addresses), then the heap reality
constraint $\heap_2 \sqsubseteq \heapg$ can be framed through $P$ if
$\heap_1$ and $\heap_2$ are separate. In other words, our frame rule
is as simple to apply as the frame rule in SL. However, to achieve
that we need to develop the concepts of heap ``enclosure'' and heap
``evolution'' and allow named subheaps to be explicitly nominated as
frames in the specifications (of functions). 

\ignore{
A significant distinction is that our frame
rule is concerned only on heap \emph{updates}, as opposed to
\emph{all} heap references as in traditional SL. Thus while SL advanced
Hoare reasoning with the implicit use of disjoint heaps, our logic
advances SL with the explicit use of arbitrary number subheaps, among
which some are disjoint and some are not. This leads to the fact that
while we cannot massage an unrestricted graph into the form $\phi
~\hsep~ \pi$ so that the SL frame rule is applicable, we \emph{can}
appropriately (and recursively) define the subheaps of the graphs so that
our new frame rule is applicable
(as demonstrated in Section ~\ref{sec:drive}).
}

\ignore{

Our new frame rule is used by explicitly
naming \emph{subheaps} in the specifications as part of the frame, in
order to elegantly isolate relevant portions of the global heap
$\cal{M}$.  Consequently, a significant distinction is that our frame
rule is concerned only on heap \emph{updates}, as opposed to
\emph{all} heap references as in traditional SL.

More specifically, we firstly facilitate the propagation of subheap
properties from the precondition to the postcondition, 
when they are not involved in program heap updates.  
This is intuitively the key
intention of a frame rule: the propagation of unaffected properties.
Secondly and just as importantly, the rule needs also to
propagate \emph{separation} information.  Toward this end, 
we introduce a concept of \emph{evolution} in a triple: when a collection of
subheaps in the precondition evolves to another collection of subheaps
in the postcondition, it follows that separation from the first
collection implies separation from the second.  Thus while SL advanced
Hoare reasoning with the implicit use of disjoint heaps, our logic
advances SL with the explicit use of arbitrary subheaps. 

}

Finally, we give evidence that our verification framework has a good
level of automation.  In Section~\ref{sec:drive}, we automatically
prove one significant example for the first time: marking a
graph. This example exhibits important relationships between data
structures that have so far not been addressed by automatic
verification: processing recursive data structures with sharing.  We
will present an implementation in Section \ref{sec:imp}, submitted as
supplementary material for this paper, and a demonstration of
automatic verification on a number of representative programs.  We
demonstrate the phases of specification, verification condition
generation and finally theorem-proving.  We stress here that we shall
be using \emph{existing} and not custom technology for the
theorem-proving.  


\ignore{
In summary, we contend that a new large of applications
is now automatically verifiable in the sense that there is an
expressive assertion language of heaps, there is a symbolic
execution algorithm to generate verification conditions from annotated
programs, and finally, these conditions can be dispatched by standard
techniques of unfolding recursive definitions.
}

\ignore{
Before proceeding, let us detail why the traditional frame rule from SL 
cannot be simply adapted to our new specification language with explicit heaps.
A first reason is explained in \cite{duck13heaps}: that with
a \emph{strongest postcondition} approach to program verification, the
frame rule, suitably translated into the language of explicit heaps,
is simply \emph{not valid}.  In other words, if $\oldtriple{\phi}{P}{\psi}$
is established because $\psi$ follows from the strongest postcondition
of $P$ executed from $\phi$, it is not the case that any heap separate
from $\phi$ remains unchanged by the execution of $P$.
A second reason is that while the assertion language refers to multiple
heaps, only those which are affected by the program must be isolated.
In contrast, the traditional rule deals with a single (implicit) heap
and so separation refers unambiguously to this heap alone.  
}

\floatstyle{boxed}
\restylefloat{figure}

\ignore{

%

At this point, note there are \emph{two kinds of footprints} at play.
One concerns what is associated with the specification describing some
data structure properties, called {\em specification footprint}; and
the other, that is concerned with the heap updates in the code or
simply the {\em code footprint}.  The key issue is how
to \emph{connect} these two in the verification process, so that
framing can take place.  As mentioned above, SL admirably addresses
these two footprints, and their connections, and there SL facilitates
the important methodology of \emph{local reasoning}.
}

\ignore{

After SL, we have the method of \emph{dynamic frames}
\cite{kassios06dynamic} (DF), and later, the refinement to
\emph{implicit} dynamic frames (IDF) ~\cite{smans09implicit}.
Essentially, a DF is a mathematical expression describing a (super)set
of memory locations that a method refers to, a.k.a. the ``footprint''
of the method.  It is used in the specification phase of verification:
together with the provided pre and postcondition of a method, the DF
represents the ``modifies'' property of the method.  The big advantage
that DF/IDF brings over SL is both in expressiveness (frames can be
specified at a higher resolution) and automation (the VC's can be
generated in a particular way so as to be able to be automatically
dispensed, perhaps with some manual intervention).

The main problem for DF is \emph{footprint compliance}:
how to navigate the code and then produce appropriate
verification conditions (VC's).
The approaches of DF and IDF require quite different answers to this,
but there is a commonality: the DF expression contains, in addition to
program variables, additional \emph{existentially quantified}
variables.  These may be (a) explicitly written as \emph{new}
variables not appearing in the program (eg. Dafny \cite{}) or (b)
implicitly due to the use of predicates with \emph{recursive
definitions}.  An example DF: a list rooted at (program variable) $x$
is either the empty set of memory locations, or \emph{there exists a
location} $next$ such that $x$ is a memory location for a pair $(val,
next)$ and DF comprises $x$ and the list rooted at $next$.  Note that
in case (b), there is an \emph{unbounded} number of existentially
quantified variables.  Hereafter we abbreviate existentially quantified
with ``ghost''.  The key to footprint compliance is then
to \emph{connect} the ghost variables to the program variables in the
method that actually traverse the data structure.  The general
approach has to \emph{annotate} the program.
{\color{magenta}
Examples ... Dafny: code on ghosts (bugs!), Viper (explicit fold/unfold)
This means some level of \emph{manual} intervention,
ie: prone to bugs and verbosity,
to connect the ghost variables to the program variables.}

We next suggest that is not sufficient just to specify a (single)
footprint of a method.
It is in fact also important to specify 
that a method is in fact, a \emph{footprint transformer}.
That is, it is important to to be able to specify that a method $M$
not only is enclosed in a particular set of memory locations,
but that it also {\color{red} Hiep ... heap summary ... etc} .
We shall demonstrate this point clearly in the next section (markgraph)

Some prominent verifiers that use DF/IDF are Vericool \cite{smans08fase},
Verifast \cite{jacobs11verifast}, Dafny \cite{leino10dafny2}, Chalice
\cite{leino09chalice} and Viper \cite{viper}.

****************************************8

1. Ghost Variables
 
* manual updates required

* verbose, buggy

* Example: inv() for append list

\vspace*{3mm}
{\color{blue}
From Kassios11tutorial: \\
Discussion. The use of ghost variables, as is supported by Dafny, is a
very flexible solution for the automation of the verification of
Dynamic Frames. The manual updates to specification-only variables
provide extremely valuable guid- ance to the prover. On the other
hand, it is an annotation overhead which may by itself be a source of
bugs. The methods prepend and prepend aux of Fig. 12 are cases where
this overhead is very heavy.  The flexibility of the Dafny language
has another price: the specifications of methods and functions become
quite verbose. This is a problem with Dynamic Frame specifications in
general, that Dafny does not address.
}

2. Fold/Unfold:

* how to control?

\vspace*{3mm}
{\color{blue} From Kassios11tutorial: \\ 

Footprint Compliance. We have already explained how footprint compli-
ance works in Chalice. The use of access predicates avoids quantified
formulas. However, it requires the programmer to explain to the
prover, in the sense of folding/unfolding how their program has
permission to write to a field.  

...

Discussion. Implicit Dynamic
Frames brings Dynamic Frames closer to Sep- aration Logic. They
notation in this theory is significantly more concise than the ghost
field solution, thanks to the use of separating conjunction. Further-
more, as we discussed above, many issues, such as footprint compliance
and self-framing checks, are performed very easily.  On the other
hand, Implicit Dynamic Frames throws away non-separating conjunction,
reverting to a form of linear logic. Such a non-standard logic, in
which specifications have side-effects to the ghost state, may be not
so intu- itive for the specifier. It is also not known how much this
loses in terms of expressiveness.  The explicit use of folds/unfolds
is very helpful for the prover, and, occa- sionally reveals
interesting bugs, but most of the time becomes tiring to the
programmer. The verifier VeriCool [39] tries to infer such statements
automati- cally, but, in the author’s opinion, for most interesting
examples this is a burden and not a help for the programmer.

}

----------------------

Our method: navigation of C GUIDES fold/unfold of F.

\vspace*{3mm}
\fbox{\color{blue} EDIT BELOW }
\vspace*{3mm}

After the development of SL, newer verification frameworks have
generally adopted the method of \emph{dynamic frames}
\cite{kassios06dynamic} (DF), and later, the refinement to
\emph{implicit} dynamic frames (IDF) ~\cite{smans09implicit}.  Some
prominent verifiers that use DF/IDF are Vericool \cite{smans08fase},
Verifast \cite{jacobs11verifast}, Dafny \cite{leino10dafny2}, Chalice
\cite{leino09chalice} and Viper \cite{viper}.  A dynamic frame is an
expression describing a set of addresses.  This set is intended to
enclose the write footprint of a method\footnote{In this context, we use 
``method'' and ``function'' interchangeably.} or code fragment.  These works
have the distinct advantage over SL: the code footprint can be
defined more precisely and further, \emph{independently} of the specification
footprint.  (Recall that in SL, the latter is used for the former.)

\ignore{
not being restricted to separating
conjunctions within the definitions of the dynamic frames (but
importantly, not within the definitions of the data structures).
Hence individual dynamic frames can
have unrestricted intersections with each other. 
} 

On the other hand, the use of dynamic frames requires additional
machinery to \emph{prove} that the heap updates (by the code) are
indeed enclosed by the appropriate dynamic frames.  (Whereas, in SL,
this is ensured by the logic itself and the accompanied inference
rules.)  For example, in some verifiers, e.g., Dafny
\cite{leino10dafny2}, ghost variables are used to explicitly describe
the dynamic frame, and the code may be annotated with ghost variable
assignments.  Correctness then requires that the heap updates are
enclosed in the distinguished ghost variable nominated as the dynamic
frame of the code.  A disadvantage is the added verbosity required on
the ghost variable expressions, and the added risk of bugs in matching
these expressions against program variable expressions.

IDF approaches, equipped with a new kind of assertion called an
\emph{accessibility predicate}, state that heap dereference
expressions (whether in assertions or in method bodies) are only
allowed if a corresponding permission has already been acquired. This
mechanism style allows a method frame to be calculated implicitly from
its precondition.  In this regard, IDF is similar to the our framework
because the accessible addresses can be contrasted with our
``enclosing'' explicit subheaps.  In particular, our notions of
``evolution'' and ``enclosure'' are closely related to the concepts of
``swinging-pivot'' and ``self-framing'' \cite{kassios11tutorial},
respectively.

However there remains a general and challenging problem that all works
using DF/IDF have not addressed: how to \emph{connect} the code
footprint (or dynamic frame) to the specification footprint when these
footprints are necessarily recursively-defined.
%
%
%
For example, it is notoriously known that many important properties of
data structures are in the form of a reachability property, and thus
they are difficult to reason about (automatically) without using
recursive definitions.  It is also known that for a large number of
programs that work on data structures, the set of nodes actually
modified by a function is a subset of what reachable from an anchor
node. Of course such a set is more naturally expressible using
recursive definitions.  Amongst the state-of-the-art verifiers, only
Vericool allows a recursive definition of its dynamic frame, and it is
generally accepted that Vericool is not an automated system.

We finally mention here, in our comparison with DF/IDF, that it is
not just the case that existing DF/IDF methods have not implemented
recursively defined specifications.  More to the point is that a
specification of a complex data structure requires an \emph{unbounded}
number of frames.  We have shown that our recursive specifications
suffice, but is not the main contribution.  Instead, it is that there
is a \emph{combination} of reasoning about the specification (again,
this must encompass an unbounded number of frames) and reasoning about code.

{\color{blue} From Muller16: \\
Despite its usefulness and inclusion in early presentations of separation logic,
no existing program verifier supports general ISCs directly. Among the tools based
on symbolic execution, Smallfoot [2] does not support ISC; VeriFast [21] and
jStar [7] allow programmers to encode some forms of ISC via abstract predicates
that can be manipulated by auxiliary operations and lemmas (in VeriFast) or
tailored rewrite rules (in jStar). For arrays, this encoding is partially supported
by libraries. However, in the general case, programmers need to provide the extra
machinery, which significantly increases the necessary manual effort. Among
the verifiers based on verification condition generation, GrassHopper [15] does
not support ISC; Chalice [11] supports only a restricted form (ranging over all
objects stored in a sequence). VeriCool uses an encoding that leads to unreliable
behaviour of the SMT solver [20, p. 46].
}

\vspace*{3mm}
\fbox{\color{blue} THERE IS HOWEVER SOME GAP}
\vspace*{3mm}

\begin{itemize}
\item Specifications \\
to describe reachability and frames simultaneously (because this is natural)
\item VC Generation \\
to have \emph{predicate transformation} (in order to have predictability in the process)
\item VC Proving \\
to accomodate induction
(again, because recursive structures are dealt with by recursive code)
\item A Frame Rule \\
(in order to trigger recursion unfolding)
\end{itemize}

\ignore{
\subsection*{Local reasoning} 
(a) EXPRESSIVENESS: because sub-programs often refer to separate heaps.
So need to explicate these heaps in the assertion language.
This is the specification footprint.
(b) FRAME RULE: to be able to COMBINE the reasoning of the sub-programs.

\subsection*{Verification Condition Generation}
2. VC generation:

(a) Strongest postcondition: to prove {P} S {Q}, we need to prove P => sp(S, P)
(b) Can do SYM EXEC for testing and BMC 

\subsection*{Verification Condition Proving}
Main contribution is here.

Now VC's are generated from two things: the assertions, and the code.
(Example: to prove {P} S {Q}, we prove P => sp(S, P) where the conclusion
is derived directly from the code S.)
In order to prove, there must be some relationship between the ``code footprint''
}

\ignore{  USE in RELATED WORK

However, there are aspects of SL which could be enhanced.
In SL, the use of predicates is overloaded: they specify
a logical property of a data structure, and at the same time,
a layout of the current heap (memory).

\begin{itemize}
\item A predicate specifying a ``large'' data structure is composed
with specifications of its constituent sub-data structures only by means of
a ``separating conjunction''.  This implies that these sub-data
structures must have disjoint footprints, an obstacle to specifying
\emph{shared} data structures.
(See \cite{hobor13ramify} for a detailed discussion of this.)

\item Predicates specify (disjoint parts of) the current heap only.
They do not connect to, e.g. previous or future heaps.
Thus, for example, it is problematic to specify a \emph{summarization}
of a program as a heap transformer.
 
\item The local proof of a function requires
that all heap accesses are ``enclosed'' by the footprint of
its precondition (or refer to fresh addresses), and  
the frame rule does not accommodate for the distinction between heap reads and writes.
This is in fact stronger than needed, because the function
might perform no heap writes.

%

\item SL does not easily provide 
some form of ``predicate transformation''
\cite{dijkstra76discipline}, which typically means to provide a
mechanism for computing either the weakest precondition or strongest
postcondition over loop-free and function-free program fragments.
Instead, SL depends on a number of custom inference rules whose
automation may not be easy.   
 \end{itemize}
}

\ignore{
However, because the separating conjunction is used in place of
logical conjunction, we cannot easily state or frame properties of \emph{shared} data structures.  
(See \cite{hobor13ramify} for a detailed discussion of this.)
The frame rule of SL used over local proofs
$\oldtriple{\phi}{P}{\psi}$ is in fact restricted to the condition
that the heap references in $P$ are \emph{enclosed} in the ``footprint'' of $\phi$.
This can restrict a fine-grained capture of the set of heap accesses of $P$.  
Further, SL does not readily support a notion of
\emph{predicate transformation}; in particular, it
does not prescribe how to compute the strongest postcondition of an
assertion.  This complicates \emph{automating} SL proofs. 
}
 
In this paper, we begin with an assertion language in which subheaps
may be \emph{explicitly} defined within predicates \cite{duck13heaps},
and the effect of separation obtained by specifying that certain heaps
are disjoint.  In other words, heaps are \emph{first-class} in this
language. One main contribution of \cite{duck13heaps} is to refine
the ``overloaded meaning'' of the separation conjunction, so that 
predicates can be conjoined in the traditional way.
In this paper, we first extend
the assertion language of  \cite{duck13heaps} by removing 
the \emph{implicit} ``heap reality'' of any subheaps
appearing in a recursive predicate. Instead, heap reality is
explicitly specified by connecting (ghost) subheaps to the
distinguished heap variable $\cal{M}$, which represents the
global heap memory at the current state. We then show how to capture
complex properties about \emph{both} sharing and separation.

Our verification framework consists of two parts.  
In the first part, we deal
with the part of a heap that is \emph{possibly changed} by a
straight-line program fragment.  This is handled by a \emph{strongest
postcondition} transform, so that the proof of a triple
$\oldtriple{\phi}{P}{\psi}$ will just require the proof of $\psi$
given the strongest postcondition of $P$ from $\phi$.  The
transformation, inherited from \cite{duck13heaps}, can be easily automated,
providing a basis towards automated verification.

\ignore{
\begin{wrapfigure}{r}{0.5\textwidth}
$\infer{\oldtriple{\phi ~\hsep~ \pi}{P}{\psi ~\hsep~ \pi}}{\oldtriple{\phi}{P}{\psi}}$ \hfill (1.1)
\end{wrapfigure}
}

Our contribution lies in the second part of the verification framework:
to perform compositional reasoning by automatically framing
properties of heap that are \emph{definitely unchanged}. 
Indeed, the main contribution of this paper is a new frame rule
to reclaim the power of local reasoning.
Before proceeding, let us detail why the traditional frame rule from SL 
cannot be simply adapted to our new specification language with explicit heaps.
A first reason is explained in \cite{duck13heaps}: that with
a \emph{strongest postcondition} approach to program verification, the
frame rule, suitably translated into the language of explicit heaps,
is simply \emph{not valid}.  In other words, if $\oldtriple{\phi}{P}{\psi}$
is established because $\psi$ follows from the strongest postcondition
of $P$ executed from $\phi$, it is not the case that any heap separate
from $\phi$ remains unchanged by the execution of $P$.
A second reason is that while the assertion language refers to multiple
heaps, only those which are affected by the program must be isolated.
In contrast, the traditional rule deals with a single (implicit) heap
and so separation refers unambiguously to this heap alone.  

Our new frame rule is used by explicitly naming \emph{subheaps} in the
specifications as part of the frame, in order to elegantly isolate
relevant portions of the global heap $\cal{M}$.   Consequently, 
a significant distinction is that our frame rule
is concerned only on heap \emph{updates}, as opposed to
\emph{all} heap references as in traditional SL.

More specifically, we firstly facilitate the propagation of subheap
properties from the precondition to the postcondition, 
when they are not involved in program heap updates.  
This is intuitively the key
intention of a frame rule: the propagation of unaffected properties.
Secondly and just as importantly, the rule needs also to
propagate \emph{separation} information.  Toward this end, 
we introduce a concept of \emph{evolution} in a triple: when a collection of
subheaps in the precondition evolves to another collection of subheaps
in the postcondition, it follows that separation from the first
collection implies separation from the second.  Thus while SL advanced
Hoare reasoning with the implicit use of disjoint heaps, our logic
advances SL with the explicit use of arbitrary subheaps.  

Finally, we give evidence that our verification framework has a good
level of automation.  In Section~\ref{sec:drive}, we automatically
prove one significant example for the first time: marking a
graph. This example exhibits important relationships between data
structures that have so far not been addressed by automatic
verification: processing recursive data structures with sharing.  We
will present an implementation in Section \ref{sec:imp}, submitted as
supplementary material for this paper, and a demonstration of
automatic verification on a number of representative programs.  We
demonstrate the phases of specification, verification condition
generation and finally theorem-proving.  We stress here that we shall
be using \emph{existing} and not custom technology for the
theorem-proving.  

In summary, we contend that a new large of applications
is now automatically verifiable in the sense that there is an
expressive assertion language of heaps, there is a symbolic
execution algorithm to generate verification conditions from annotated
programs, and finally, these conditions can be dispatched by standard
techniques of unfolding recursive definitions.

 

\ignore{ 456
Another shortcoming concerns automation.
As explained in \cite{duck13heaps}: suppose we interpret
triples as in classic Hoare-logic after having extending the
assertion language to accommodate the expression of subheaps.
The frame rule, suitably translated into this language of subheaps,
is simply not valid (without some additional machinery ensuring enclosure).  
In other words, if $\oldtriple{\phi}{P}{\psi}$
is established because $\psi$ follows from the \emph{strongest postcondition}
of $P$ executed from $\phi$, it is not the case that any subheap separate
from $\phi$ remains unchanged by the execution of $P$.
Another aspect is that while the assertion language refers to multiple
heaps, only those which are affected by the program must be isolated.
In contrast, the traditional rule deals with a single (implicit) heap
and so separation refers unambiguously to this heap alone.  
These shortcomings represent the motivation for this paper
to introduce a new program verification framework which 
\begin{itemize}
\item
accommodates local reasoning,
\item 
expressive
enough to accommodate general data structures (and in particular, shared
data structures), and which is 
\item
based on traditional strongest postcondition
propagation methods in conjunction with traditional Hoare logic
in order to achieve a good level of automation.
\end{itemize}

\noindent
At the center of our framework is a new frame rule which accommodates heap sharing between
assertion predicates (e.g. when defining graphs), with the crucial
property that it supports \emph{automatic verification}: given an annotated program,
our frame rule generates verification conditions (VCs) which can
be dispensed by an appropriate theorem-prover.
In contrast, the frame rule of SL is not always applicable,
because it is preconditioned on the fact that the footprints of 
relevant program fragments are disjoint. 
There have been works, e.g. \cite{hobor13ramify}, to widen the applicability
of compositional reasoning, but here the VC's generated are complex (e.g. they
contain the magic-wand operation from SL) and thus cannot be plausibly
dispensed by a theorem-prover automatically. 


We begin with an existing assertion language
(in Section~\ref{sec:heapformulas}) in which subheaps may
be \emph{explicitly} defined within predicates \cite{duck13heaps}, and
the effect of separation obtained by specifying that certain heaps are
disjoint.  Thus heaps are \emph{first-class} here.  We also borrow
from \cite{duck13heaps} the \emph{strongest postcondition} transform
(in Section~\ref{sec:symexec}) so that 
proofs of programs without loops or procedure calls are automated up
to the use of a (heap-aware) theorem-prover.  We however make
important additions to the program verification framework
of \cite{duck13heaps} by adding the notion of \emph{ghost subheaps},
and this provides the basis for the expression of unrestricted
sharing.  We then follow this up with a notion of \emph{heap reality}.
Finally, we present a number of new rules, including a new frame rule,
so as to obtain a new program verification framework.  In the end, our
main technical contribution over \cite{duck13heaps} is the provision
of local reasoning.

\ignore{
Next, we present the verification framework in two parts.  First, we deal
with the part of a heap that is \emph{possibly changed} by a
straight-line program fragment.  This is handled by a \emph{strongest
postcondition} transform, so that the proof of a triple
$\oldtriple{\phi}{P}{\psi}$ will just require the proof of $\psi$
given the strongest postcondition of $P$ from $\phi$.  This
transformation, inherited from \cite{duck13heaps}, has been fully automated,
providing a first basis toward automated verification.
}

\ignore{
\begin{wrapfigure}{r}{0.5\textwidth}
$\infer{\oldtriple{\phi ~\hsep~ \pi}{P}{\psi ~\hsep~ \pi}}{\oldtriple{\phi}{P}{\psi}}$ \hfill (1.1)
\end{wrapfigure}
}

Our new rule (in Section~\ref{sec:framerule})
is used by explicitly naming \emph{subheaps} in the
specifications as part of the frame, in order to elegantly isolate
relevant portions of the global heap.   Consequently, 
a significant distinction is that our frame rule
is concerned only on heap \emph{updates}, as opposed to
\emph{all} heap references as in traditional SL.
More specifically, we firstly facilitate the framing of subheap properties 
over a program fragment when they are not involved in program heap updates. 
This is intuitively the key intention of a frame rule:
the propagation of properties of unaffected parts of the heap.
Secondly and just as importantly, the rule also needs  to
propagate \emph{separation} information, which is crucial
for successive uses of framing.  Toward this end, 
we introduce a concept of \emph{evolution} in a triple: when a collection of
subheaps in the precondition evolves to another 
in the postcondition, it follows that separation from the first
collection implies separation from the second.  Thus while SL advanced
Hoare reasoning with the implicit use of disjoint heaps, our logic
advances SL with the explicit use of arbitrary subheaps.  
} 

}

\section{A Challenge Problem}
\label{sec:challenge}

In Fig.~\ref{fig:markgraph} we have a textbook algorithm that marks a 
graph.  We assume a node has two successor fields \texttt{left} and \texttt{right}.  
\ignore{
Note that because the {\tt left} subgraph and the {\tt right} subgraph may overlap, we
cannot (directly) apply SL.  Instead, we shall demonstrate what it
would take to prove this program, with emphasis on roles of the
specification and code footprints. 
}

\begin{figure}[tbh]
\begin{center}
\small{
\stuff{
struct node \{ int m; struct node $*$left, $*$right; \}; \\
\\
void mark(struct node $*$x) \{ \\
\> if (!x || x->m) return; \\
\>   struct node $*$l = x->left, $*$r = x->right; \\
\>    x->m = 1; mark(l); mark(r); \\
\}
}
}
\end{center}
\caption{Mark Graph Example}
\label{fig:markgraph}
\end{figure}

\ignore{
The top-level precondition is that the graph is unmarked, and the end
result is that the graph is fully marked. Because the function is
recursive, clearly its precondition cannot simply be that the graph is
fully unmarked.  The required precondition is rather complicated, and
we relegate the details to Section \ref{sec:drive}.  Here it suffices
to say that the precondition must state that every encountered marked
node is either previously encountered, or all of its successor nodes
are already marked.  The take-away is that this property is not
naturally expressible without using a recursive definition.
}

There are some subtle but critical points that makes the example
extremely challenging.  First, in order to have a well-founded recursive
definition of a graph, we need some form of ``history''.
Yet the program itself \emph{does not implement} any
such notion.    Instead it uses the mark ({\tt m}) field for termination.  
Thus a challenge is to \emph{connect} a node's history 
in the specification and its mark.

\begin{wrapfigure}{r}{0.2\textwidth}
\vspace{-2mm}
\begin{center}
\includegraphics[width=0.19\textwidth]{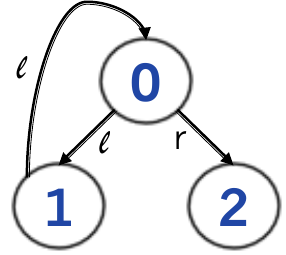}
\caption{A Cyclic Graph}
\label{fig:graph}
\end{center}
\vspace{-2mm}
\end{wrapfigure}

A second critical point is: what is the \emph{precise specification}
of the function?  It is clear that we eventually want a fully marked
graph, but because the program is recursive, the pre and postcondition
also act as {\em invariants} and are somewhat complicated.  Clearly
the precondition cannot be an arbitrary graph, otherwise the program
may \emph{prematurely} terminate.  One intuitive precondition is that
the graph is ``mark successor closed'', i.e., any successor node of a
marked node is itself marked.  This concept covers both fully unmarked
graphs as well as fully marked graphs.  However, this intuitively
appealing condition is, surprisingly, \emph{too strong}.  To prove
this, consider the initially unmarked graph in Fig.~\ref{fig:graph}
and we start the {\tt mark} function at the root node $0$.  During
processing, at the start of the second visit to node $0$ (by going 
\texttt{left} and then \texttt{left} again) the graph no longer satisfies the
proposed precondition because while $0$ has been marked, one of its
successors, node $2$, has not yet been marked.

The actual required precondition is rather complicated, and we relegate the
details to Section \ref{sec:drive}.  Here it suffices to say that the
precondition must state that every encountered marked node is either
previously encountered, or all of its successor nodes are already
marked.  The take-away is that this property is not naturally
expressible without using a recursive definition.


The third critical point is to achieve local reasoning.
We must ensure that the first recursive call 
must not negate what is needed as the
precondition of the second call 
and dually the second call should not negate the effects of the first.  
In other words, we need to describe: (1) the write footprint of the first
call, (2) the footprint of the precondition of the second call, (3)
the footprint of the postcondition of the first call, and (4) the
write footprint of the second call. The verification process needs to
automatically figure out that (1) and (2) are disjoint and that (3)
and (4) are also disjoint.
  

\ignore{
\hcomment{HIEP: the following sentence is unclear or broken} 
But perhaps more importantly, this example displays a formal
of a basic algorithm which traverses a complex data structure and is
able to connect the formal specification, written in the standard form
of a well-founded recursive definition, and the code, which operates
using the standard method of marking.  We elaborate in
Section \ref{sec:drive}.
}


\ignore{
here means the
following: given an annotated program, where there are assertions
appropriately placed, that (I) we can generate appropriate VC's, and (II) we can
employ a systematic algorithm to solve the VC's.
\fbox{expand}

Now why is this problem (I) and (II) especially challenging, for
example, as opposed to programs which manipulate data structures
which need not be regarded or treated as recursive?
}

\ignore{

Reconsider the graph marking example, whose program was presented earlier in
Fig.~\ref{fig:markgraph}.  Now, initially the graph is unmarked, and
we want to prove that at the end, the graph is fully marked.  The
definition of {\tt mgraph} in Figure~\ref{fig:def:mark} simply states
that a graph is fully marked. A node is marked if its \texttt{mark}
field is 1; otherwise if the value is 0. The parameter $t^{in}$ is of
heap type, representing the ``history'' that includes all the visited
nodes --- starting off from a root node of interest and an empty history. 
The usage of a ``history'' is critical in defining a possibly
{\em cyclic} graph.

There are some subtle but critical points that makes the example
extremely challenging. First, note that despite the need for a history
in the specification, the program itself \emph{does not implement} any
such notion.  But without some form of history, {\em how does the
program ensure termination}?  The answer is, intuitively, that it 
uses the {\tt mark} field for termination.  Thus one of the central difficulties
example is in fact to make a connection between a node's history and its mark.

}


\section{The Assertion Language}
\label{sec:heapformulas}

We assume a vanilla imperative programming language with functions but
no loops (which are tacitly compiled into tail-recursive functions).
Other than standard non-heap statements, the following are \emph{heap
manipulation statements:}\footnote{We assume (de)allocation of single
heap cells; this can be easily generalized, and indeed so in our
implementation.}

\vspace*{2mm}
\begin{itemize} \itemsep 0mm
\item sets $x$ to be the value pointed to by $y$: \hfill  {\tt x = $*$y};
\item sets the value pointed to by $x$ to be $y$: \hfill  {\tt $*$x = y}; 
\item points $x$ to a freshly cell: \hfill {\tt x = $\mathbf{malloc}$(1)}; 
\item deallocates the cell pointed to by $x$: \hfill {\tt $\mathbf{free}$(x)}.
\end{itemize}
\vspace*{2mm}

\noindent
The heap is not explicitly mentioned in the program.  Instead, it is
dereferenced using the ``$\hsep$'' notation as in the C language.
(Not to be confused with the operator ``$\hsep$'' in SL or our heap
constraint language.)  Since our later discussion will involve
symbolic execution, we also assume that branch condition is translated
to $assume(\_)$ statement.  We first give a brief overview of Hoare
and Separation Logic, then we introduce the language
in~\cite{duck13heaps}.

\subsection{Background}

\emph{Hoare Logic}~\cite{hoare69axiom} 
is a formal system for reasoning about program correctness.
Hoare Logic 
is defined in terms of axioms over \emph{triples} of the
form $\oldtriple{\phi}{P}{\psi}$, where 
$\phi$ is the \emph{precondition}, $\psi$ is the \emph{postcondition},
and $P$ is some code fragment.
Both $\phi$ and $\psi$ are formulae over the \emph{program variables} in $P$.
The meaning of the triple is as follows:
for all program states $\sigma_1$, $\sigma_2$ such that
$\sigma_1 \models \phi$ and executing $\sigma_1$ through $P$ derives
$\sigma_2$, then $\sigma_2 \models \psi$.
For example,  the triple 
$\oldtriple{x < y}{{\tt x} = {\tt x} + 1}{x \leq y}$, x and y are integers, 
is \emph{valid}. Note that under this definition, a triple is automatically valid if
$P$ is non-terminating or has undefined behavior.
This is known as \emph{partial correctness}.

\emph{Separation Logic} (SL) ~\cite{reynolds02separation} is a popular extension of
Hoare Logic ~\cite{hoare69axiom} for reasoning over \emph{heap manipulating programs}.
SL extends predicate calculus with new logical 
connectives -- namely \emph{empty heap} ($\mathbf{emp}$),
\emph{singleton heap} ($p \mapsto v$), and
\emph{separating conjunction} ($F_1 \hsep F_2$) --
such that the structure of assertions reflects the structure of the
underlying heap.
For example, the precondition in the following valid Separation Logic triple

\vspace*{1mm}
\begin{center}
$\oldtriple{x \mapsto \_ ~\hsep~ y \mapsto 2}{\texttt{$*$x} = \texttt{$*$y} + 1}{x \mapsto 3 ~\hsep~ y \mapsto 2}$
\end{center}
\vspace*{1mm}

\noindent
represents a heap comprised of two \emph{disjoint singleton} heaps,
indicating that both $x$ and $y$ are \emph{allocated} and that
location $y$ points to the value $2$.
In the postcondition, $x$ points to value $3$, as expected.
SL also allows \emph{recursively-defined} heaps for reasoning
over data structures, such as $\mathbf{list}$ and
$\mathbf{tree}$.
%
An SL \emph{triple} $\oldtriple{\phi}{P}{\psi}$ additionally
guarantees that any state satisfying $\phi$ will not cause a memory
access violation in $P$.  For example, the triple
$\oldtriple{\mathbf{emp}}{\texttt{$*$x} := 1}{x \mapsto 1}$ is
\emph{invalid} since $x$ is a dangling pointer in a state satisfying the
precondition.

\vspace{2mm}
\noindent
{\bf A Constraint Language of Explicit Heaps}~\cite{duck13heaps}:
We have a set of $\set{Values}$ (e.g. integers) and 
we define $\set{Heaps}$ to be all
\emph{finite partial map}s between values, i.e.,
$\set{Heaps} \eqdef
(\set{Values} \rightharpoonup_{\text{fin}} \set{Values})$.  There is a
special value $\Null$ (``null'' pointer) and a special heap $\hemp$
(``empty'' heap). Where $\vv_v$ and $\vv_h$ denote the sets of value
and heap variables respectively, our \emph{heap expressions} $H\!E$
are as follows:

\vspace*{0mm}
\begin{center}
$\begin{array}{l}
H ::= \vv_h \hspace{10mm}
v ::= \vv_v \\
H\!E ::= H ~|~ \hemp ~|~ \hone{v}{v} ~|~ H\!E ~\hsep~ H\!E
\end{array}$
\end{center}
\vspace*{0mm}

\noindent
An \emph{interpretation} ${\cal I}$ maps $\vv_h$ to $\set{Heaps}$ and
$\vv_v$ to $\set{Values}$.  Syntactically, a \emph{heap constraint} is
of the form $(H\!E \heq H\!E)$.  An interpretation ${\cal
I}$ satisfies a heap constraint $(H\!E_1 \heq H\!E_2)$ iff $ {\cal
I}(H\!E_1) = {\cal I}(H\!E_2)$ are the same heap, and the separation
properties within $H\!E_1$ and $H\!E_2$ hold.

Let $\dom(H)$ be the \emph{domain} of the heap $H$.  As in~\cite{duck13heaps}, 
heap constraints can be normalized into three
basic forms:

\vspace*{0mm}
\begin{center}
$\begin{array}{l}
H \heq \hemp \hfill ~~~\textsc{(Empty)} \hspace{10mm}
H \heq \hone{p}{v} \hfill ~~~\textsc{(Singleton)} \\
H \heq H_1 ~\hsep~ H_2 \hfill ~~~\textsc{(Separation)}
\end{array}$
\end{center}
\vspace*{0mm}

\noindent
where $H, H_1, H_2 \in \vv_h$ and $p, v \in \vv_v$.  Here
(\textsc{Empty}) constrains $H$ to be the empty heap (i.e., $H
= \emptyset$ as a set), (\textsc{Singleton}) constrains $H$ to be the
singleton heap mapping $p$ to $v$ (i.e., $H = \{(p, v)\}$ as sets),
and (\textsc{Separation}) constraints $H$ to be the heap that is
partitioned into two disjoint sub-heaps $H_1$ and $H_2$ (i.e., $H =
H_1 \cup H_2$ as sets and $\dom(H_1) \cap \dom(H_2) = \emptyset$).

We will also use \emph{sub-heap relation}  ($H_1 \sqsubseteq H_2$), 
\emph{domain membership} ($p \in \dom(H)$),
and (overloaded) for brevity, \emph{separation relation} ($H_1 ~\hsep~ H_2$).
In fact, writing $H_1 \sqsubseteq H_2$ is equivalent to $H_2 \heq H_1 ~\hsep~ \_$,
 $p \in \dom(H)$ to $H \heq \hone{p}{\_} ~\hsep~ \_$, 
 $p \not\in \dom(H)$ to $\_ \heq H ~\hsep~ \hone{p}{\_}$,
 and $H_1 ~\hsep~ H_2$ to $\_ \heq H_1 ~\hsep~ H_2$;
where the underscore in each instance denotes a fresh variable.

Finally, we have a \emph{recursive constraint}.  This is an expression of the form 
%
$p(h_1, \cdots, h_n, v_1, \cdots , v_m)$ 
%
where $p$ is a user-defined \emph{predicate symbol},
the $h_i \in \vv_h, 0 \leq i \leq n$ and the $v_j \in \vv_v, 0 \leq j \leq m$.
Associated with such a predicate symbol is a \emph{recursive definition}.
We use the framework of \emph{Constraint Logic Programming} 
(CLP)~\cite{jaffar87clp} to inherit its syntax, semantics, and its
built-in notions of unfolding rules, for realizing recursive definitions.
The \emph{semantics} of a set of rules is traditionally 
known as the ``least model'' semantics \cite{jaffar87clp}.
For brevity, we only informally explain the language.
The following constitutes a recursive definition of $\texttt{list}(h, x)$,
specifying a \emph{skeleton list} in the heap $h$ rooted at $x$.

\vspace*{2mm}
\stuff{
list($h, x$) \clpif~ $h \heq \hemp, x = \Null$. \\
list($h, x$) \clpif~ $h \heq \hone{x}{y} ~\hsep~ h_1$, list($h_1, y$).
}

\vspace*{2mm}
\noindent
Note that the comma-separated expressions in the body of each rule is
either \emph{value constraint} (e.g. $x = \Null$), a heap
constraint (e.g.  $h \heq \hemp$), or a recursive constraint
(e.g. \texttt{list($h_1, y$)}).  In this paper,
our value (i.e. ``pure'') constraints will
either be arithmetic or basic set constraints over values.


\subsection{Program Verification with Explicit Heaps}

\noindent
{\bf Hoare Triples:} We first define an \emph{assertion} $A$ as a
formula over $\vv_v$, $\vv_h$:

\vspace{0mm}
\begin{center}
$A ::= V\!F ~~|~~ H\!F ~~|~~ RC ~~|~~ A \wedge A ~~|~~ A \vee A $
\end{center}
\vspace{0mm}

\noindent
where $V\!F$, $H\!F$, and $RC$ are value, heap, and recursive constraints, respectively.
We next connect the interpretation of assertions with the program semantics.

Programs operate over an unbounded set of \emph{program
variables} ${\cal V}_{\!P}$, which are the \emph{value variables}.
Thus ${\cal V}_{\!P} \subseteq \vv_v$. We use one distinguished heap
variable $\heapg \in \vv_h$ to represent the \emph{global heap
memory}. Variables other than the program variables and $\heapg$ may
appear in assertions; they are existential or \emph{ghost}
variables. A ghost variable of type heap will be called
a \emph{subheap}.

The subheaps serve two essential and distinct purposes: (a) to
describe subheaps of the global heap $\heapg$ at the current program
point, and (b) to describe some other ``existential'' heap.  A common
instance of (b) is the heap corresponding to the global heap at
some \emph{other} program point in the past.

We use the terminology ``ghost heap'' in accordance to standard practice
that subheaps are existential, but in assertions, they can be used to
constrain the value of the global heap.  Importantly, as ghost variables,
their values \emph{cannot be changed} by the program.
We will see later that this is important in practice because (a) predicates
in assertions often need to be defined only using ghost subheaps, and (b)
it is automatic that these predicates can be ``framed through'' any program fragment $P$
because $P$ cannot change the value of a ghost variable.

\ignore{
An \emph{interpretation} of an assertion is obtained in the traditional way:
value terms are interpreted into values, and heap terms are interpreted into finite
maps from values to values, and finally value and heap formulas are interpreted
into \emph{false} or \emph{true} (in which case it is a \emph{model}).}

Note that we shall present rules that define recursive constraints
using fresh variables.  Notationally, for heaps, we shall use the
small letter `h' in rules, while using the large letter $\heap$ in
assertions. Also, we use ``\texttt{,}'' in assertions as shorthand for
logical conjunction.

\vspace{1mm}
\noindent
{\bf Example:} see the annotated program and the definition of {\tt inc\_list}
in Fig.~\ref{fig:ex:inc}.
The program increments all the data values in an acyclic list by 1.   


\begin{figure}[h!]
\small{
\stuff{
struct node \{ int data; struct node $*$next; ~\}; \\
\ \\

\texttt{inc\_list}($h_1, h_2, x$) \clpif ~$h_1 \heq \hemp, h_2 \heq \hemp, x = \Null$. \\
\texttt{inc\_list}($h_1, h_2, x$) \clpif ~$h_1 \heq \hone{x}{(d+1, next)} * h_1'$, \\
\> $h_2 \heq \hone{x}{(d, next)} * h_2'$, \texttt{inc\_list}($h1', ~h_2', ~next$). \\
}
\ \\
\hspace*{-17mm}\begin{minipage}{60mm}
\stuff{
$\hiepassert{\{~list(\heap, x) \hiepand \heap \sqsubseteq \heapg~\}}$ \\
\> y = x; while (y) \{ y->data += 1; y = y->next; \} \\
$\hiepassert{\{~inc\_list(\heap_1, \heap, x) \hiepand \heap_1 \sqsubseteq \heapg~\} } $ 
}
\end{minipage}
}
\caption{Incrementing data values in an acyclic list.}
\label{fig:ex:inc}
\end{figure}

The recursive constraint \texttt{list}($\heap, x$) describes a heap $\heap$ which
houses an acyclic list rooted at $x$.  The constraint
$\heap \sqsubseteq \heapg$ states that it resembles a part of the
global heap.
The other recursive constraint 
\texttt{inc\_list}($\heap_1, \heap, x$) 
similarly defines that $x$ is the head of a list resides in the heap $\heap_1$.
It has another argument, the ghost heap $\heap$, which also appears in
the precondition.  This, importantly, allows us to consider the triple
as a \emph{summary}, relating values in the precondition and
postcondition (using the ghost variable as an anchor value).  In this
case, we are stating that the final list elements are one bigger than
the corresponding initial elements.  Further, we are also stating that
all the links (the $next$ pointers) are not modified.

\ignore{
Note that in a triple, a ghost variable appearing
\emph{both} in the precondition and postcondition means that we are referring
to the same entity,
e.g., $\heap$ above, while this is not the case for the program variable $x$ or for
the global heap memory $\heapg$.
(By analogy, in Hoare Logic the 
triple \{$x = x_0$\} \texttt{x++} \{$x = x_0+1$\} describes an increment of $x$.)

of changes of subheaps done by the code
The definition of \texttt{inc\_list}, shown above,
not only states that each datum in the list rooted at $x$ in $\heap_1$  
is one more that the corresponding datum in the list rooted at $x$ in $\heap$,
but also states that all the links (the $next$ pointers) are not modified.
}

\ignore{
Put another way, in the postcondition,
(a) the new heap of interest housing the list in the memory 
is $\heap_1$ (by virtue of the
constraint $\heap_1 \sqsubseteq \heapg$), and
(b) the values in this list are related to the list in the precondition because
the latter list has been captured in the variable $\heap$, 
which in the postcondition is a ghost variable. It is important to note also that
the recursive constraint $\texttt{list}(\heap,x)$ is unaffected by the code and
still holds at the postcondition, though its \emph{usefulness} is questionable. 
On the other hand, $\heap \sqsubseteq \heapg$
no longer holds at the postcondition, because the memory $\heapg$ has been updated.
}

\ignore{
\noindent
A second example shows how subheap variables can be used to specify 
\emph{different} parts of the global heap.
This is a classic example demonstrating the succinctness of SL, and our purpose here 
is simply to show that our setting shares this advantage.

We start with a definition of binary tree whose only data value is a mark.
Below we define the predicate \texttt{tree}($\heap, x$) where $\heap$ is a heap
containing a tree of nodes rooted at $x$.
Each node is a 3-element heap comprising a mark field, and two other fields
for the left and right successor nodes of $x$.
We also define a similar predicate \texttt{markedtree}($h, x$) which
is similar to \texttt{tree}($h, x$) except that in this case,
all the mark fields in the tree have the value 1.

\vspace*{3mm}
\stuff{
tree($h, x$) \clpif~$h \heq \hemp, x = \Null$. \\
tree($h, x$) \clpif  \\
\> $h \heq (x \mapsto (\_, \lef, right)) * h_1 * h_2$, \\
\> tree($h_1, \lef$), \\
\> tree($h_2, right$). \\
\\
markedtree($h, x$) \clpif~$h \heq \hemp, x = \Null$. \\
markedtree($h, x$) \clpif  \\
\> $h \heq (x \mapsto (1, \lef, right)) * h_1 * h_2$, \\
\> markedtree($h_1, \lef$), \\
\> markedtree($h_2, right$).
}
\vspace*{3mm}

\noindent
Now consider a classic algorithm for marking trees.
We use the C program below for this:

\vspace*{3mm}
\stuff{
struct node \{ \\
\> int m; \\
\> struct node *left, *right; \\
\}; \\
\ \\ 
{\bf requires:} tree($\heap, x$) $\wedge~ \heap \sqsubseteq \heapg$ \\
{\bf ensures:} markedtree($\heap', x$)  $\wedge~ \heap' \sqsubseteq \heapg$ \\
void mark(struct node *x) \{ \\
\> if (!x) return; \\
\> \hiepassert{\{ $(x \mapsto (\_, \lef, right)) * \heap_1 * \heap_2 \sqsubseteq \heapg~\wedge$ } \\
\> \hiepassert{tree($\heap_1, \lef$) $\wedge$ tree($\heap_2, right$) \} } \\
\> x->m = 1; \\
\> \hiepassert{\{ $(x \mapsto (1, \lef, right)) * \heap_1 * \heap_2 \sqsubseteq \heapg~\wedge$ } \\
\> \hiepassert{tree($\heap_1, \lef$) $\wedge$ tree($\heap_2, right$) \} } \\
\> mark(x->left); \\
\> \hiepassert{\{ $(x \mapsto (1, \lef, right)) * \heap'_1 * \heap_2 \sqsubseteq \heapg~\wedge$ } \\
\> \hiepassert{markedtree($\heap'_1, \lef$) $\wedge$ tree($\heap_2, right$) \} } \\
\> mark(x->right); \\
\> \hiepassert{\{ $(x \mapsto (1, \lef, right)) * \heap'_1 * \heap'_2 \sqsubseteq \heapg~\wedge$ } \\
\> \hiepassert{markedtree($\heap'_1, \lef$) $\wedge$ markedtree($\heap'_2, right$) \} } \\
\}
}
\vspace*{3mm}

\note{Refine these steps}

\hiepalert{

\noindent
The proof of the Hoare triple may be outlined as follows.
In this description, we shall limit our attention to 
(a) obtaining an assertion as the postcondition of the first recursive call,
and (b) \emph{framing} this assertion through the second call.

\begin{itemize}
\item When $x$ is \Null, the proof can be trivially established.

\item Assume \texttt{tree}($\heap, x$) holds for some $\heap$ that is part
of the global heap memory $\heapg$, and that $x$ is not \Null.
Going to the {\bf else} body of the {\bf if} statement, we unfold 
the predicate using second rule in the definition of \texttt{tree},
obtaining

\begin{center}
$\{~ \heap = (x \mapsto (1, \lef, right)) * \heap_1 * \heap_2 ~\wedge$ \\
\texttt{tree}($\heap_1, \lef$) $\wedge$ \texttt{tree}($\heap_2, right$) \}.
\end{center}

Note in particular that $\heap_1 * \heap_2$ holds.
\item After the first recursive call, it is established, by induction, that 
since \texttt{tree}($h_1, \lef$) holds before the call, \texttt{markedtree}($h_1, \lef$) holds
after the call.   
\item Similarly, the second recursive call establishes that \\
\texttt{markedtree}($h_2, right$) holds after the call.
\item Now in order to prove that \texttt{markedtree}($h_1, \lef$) 
\emph{continues to hold} after the second recursive call, we would use a frame rule.
\item In our setting, the (classic) frame rule 
applies because in the second call, all heap updates
are limited to (the domain of) $h_2$ which is not in the 
footprint of \texttt{markedtree}($h_1, \lef$).
This is the critical step we would like to highlight, for further reference
below.
\item Finally, that the root node $x$ is marked (in $h$) and that 
the two subtrees of $x$ are marked
will show, using an appropriate theorem-prover, 
that the entire tree rooted at $x$ is marked.
(We will comment on the theorem-proving of our assertions later.)
\end{itemize}
}

}

We conclude this section by stressing that our \emph{interpretation} of 
triples follows Hoare logic: the postcondition holds
\emph{provided} the start state satisfies the precondition, 
\emph{and} there is a terminating execution of the
program. In contrast, in SL, a triple entails that the program is
memory-safe. Though we do not provide for memory safety as an
intrinsic property, we can easily enforce memory safety, e.g., by
asserting that dereferences (e.g., \texttt{x->next}) and deallocations
(e.g., \texttt{free(x)}), have their arguments ({\tt x}) pointing to a
valid cell in the global heap ($x \in dom({\cal M})$).  Not
enforcing memory safety up front is not a weakness. It allows us to be
flexible enough to perform reasoning even when memory safety is not
the property of interest. Furthermore, SL may disapprove of a memory
safe program whose specifications of some functions are not
sufficiently complete. In contrast, our framework can still proceed,
but possibly not by means of local reasoning.

\section{Symbolic Execution with Explicit Heaps}
\label{sec:symexec}
Symbolic execution of a program uses
\emph{symbolic values} as inputs, and 
can be used for program verification in a standard way.
We start with a precondition.
The output of symbolic execution on a program path
is a formula representing the symbolic state 
obtained at the end of a path, or the \emph{strongest postcondition}
of the precondition.  For a loop-free program with no function calls, 
symbolic execution facilitates verification by considering a disjunction of all such
path postconditions, which must then imply the desired postcondition.
With function calls (or loops), to achieve modular verification, we
need a frame rule.

We now describe how to obtain a 
the strongest postcondition transform as in \cite{duck13heaps}.
It suffices to consider only the four heap-manipulating primitives.

\begin{proposition}[Strongest Postcondition]\label{prop:exe}
In the following Hoare-triples, the postcondition shown
is the strongest postcondition of the primitive heap operation
with respect to a precondition $\phi$. 

\vspace{4mm}
\hspace*{-4mm}$\begin{array}{ll}
\oldtriple{~\phi~}{x = \mathbf{malloc}(1)}{~\mathsf{alloc}(\phi, x)~} &
    (\text{Heap allocation}) \vspace*{1mm}
\\
\oldtriple{~\phi~}{\mathbf{free}(x)}{~\mathsf{free}(\phi, x)~} & 
    (\text{Heap deallocation}) \vspace*{1mm}
\\
\oldtriple{~\phi~}{x = *y}{~\mathsf{access}(\phi, y, x)~} & 
    (\text{Heap access}) \vspace*{1mm}
\\
\oldtriple{~\phi~}{*\!x = y}{~\mathsf{assign}(\phi, x, y)~} & 
    (\text{Heap assignment}) 
\end{array}$

\vspace{2mm}
\noindent
where the auxiliary macros 
$\mathsf{alloc}$, $\mathsf{free}$, $\mathsf{access}$, and $\mathsf{assign}$ 
expand as follows: \\
\vspace*{3mm}
$\begin{array}{lll}
\mathsf{alloc}(\phi, x)  & \eqdef &
	\heapg \heq \hone{x}{v} ~\hsep~ \heap \land
    \phi[\heap/\heapg, v_1/x] \vspace*{0mm} \\
\mathsf{free}(\phi, x)  & \eqdef &
	\heap \heq \hone{x}{v} ~\hsep~ \heapg \land
    \phi[\heap/\heapg] \vspace*{0mm} \\
\mathsf{access}(\phi, y, x)  & \eqdef &
	 \heapg \heq \hone{y}{x} ~\hsep~ \heap \land \phi[v/x] \vspace*{0mm} \\
\mathsf{assign}(\phi, x, y)  & \eqdef & 
	\heapg \heq \hone{x}{y} ~\hsep~ \heap_1 \land \\
    & & \heap \heq \hone{x}{v} ~\hsep~ \heap_1 \land \phi[\heap/\heapg] 
\end{array}$
\vspace*{1mm}

\noindent
where $\heap$ and $\heap_1$ are \emph{fresh} heap variables,
and $v$ and $v_1$ are fresh value variables.
The notation $\phi[x/y]$ means formula $\phi$ with variable $x$
substituted for $y$.
$\Box$
\label{thm:sp}
\end{proposition}

\ignore{
\begin{proposition}[Strongest Postcondition]\label{prop:exe}
In the following Hoare-triples, the postcondition shown
is the strongest postcondition of the primitive heap operation
with respect to a precondition $\phi$. 

\vspace{2mm}
\hspace*{-7mm}$\begin{array}{ll}
\oldtriple{~\phi~}{x = \mathbf{malloc}(1)}{~\heapg \heq \hone{x}{v} ~~\hsep~~ \heap \land
    \phi[\heap/\heapg, v_1/x]~} &  \\
\oldtriple{~\phi~}{\mathbf{free}(x)}{~\heap \heq \hone{x}{v} ~\hsep~ \heapg \land
    \phi[\heap/\heapg]~} & \\
\oldtriple{~\phi~}{x = *y}{~\heapg \heq \hone{y}{x} ~\hsep~ \heap \land \phi[v/x]~} & \\
\oldtriple{~\phi~}{\!*\!x = y}{~\heapg \heq \hone{x}{y} ~\hsep~ \heap_1 \land 
 \heap \heq \hone{x}{v} ~\hsep~ \heap_1 \land \phi[\heap/\heapg]~} & 
\end{array}$

\vspace{2mm}
\noindent
where $\heap$ and $\heap_1$ are \emph{fresh} heap variables,
and $v$ and $v_1$ are fresh value variables.
The notation $\phi[x/y]$ means formula $\phi$ with variable $x$
substituted for $y$.
$\Box$
\label{thm:sp}
\end{proposition}
}

\begin{figure}[h!]
\begin{center}
\stuff{
\hiepassert{$\{ ~\heap_{99} \heq \heapg ~\}$} \\
\> t$_1$ = $*$x; \\
\hiepassert{$\{ ~\heapg \heq \hone{x}{t_1} ~\hsep~ \heap_1 \hiepand \heap_{99} \heq \heapg ~\}$} \\
\> $*$x = t$_1$ + 1; \\
\hiepassert{$\{ ~\heapg \heq \hone{x}{t_1+1} ~\hsep~ \heap_1 \hiepand \heap_2 \heq \hone{x}{t_1} ~\hsep~ \heap_1 \hiepand$} \\
\hiepassert{$~~\heap_2 \heq \hone{x}{t_1} ~\hsep~ \heap_1 \hiepand   \heap_{99} \heq \heap_2  ~\}$} \\
\vspace{2mm}
\> $\Downarrow$ ~//  (simplification)\\ 
\vspace{2mm}
\hspace{-2mm}\hiepalert{$\{ ~\heapg \heq \hone{x}{t_1+1} ~\hsep~ \heap_1 \hiepand \heap_{99} \heq \hone{x}{t_1} ~\hsep~ \heap_1 ~\}$}  \\
\> t$_2$ = $*$x; \\
\hiepassert{$\{ ~\heapg \heq \hone{x}{t_2} ~\hsep~ \heap_3 \hiepand \heapg \heq \hone{x}{t_1+1} ~\hsep~ \heap_1  \hiepand $} \\
\hiepassert{$~~\heap_{99} \heq \hone{x}{t_1} ~\hsep~ \heap_1 ~\}$} \\
\> $*$x = t$_2$ - 1; \\
\hiepassert{$\{ ~\heapg \heq \hone{x}{t_2-1} ~\hsep~ \heap_4 \hiepand \heap_5 \heq \hone{x}{v} ~\hsep~ \heap_4  \hiepand $} \\
\hiepassert{$~~\heap_5 \heq \hone{x}{t_2} ~\hsep~ \heap_3 \hiepand \heap_5 \heq \hone{x}{t_1+1} ~\hsep~ \heap_1  \hiepand$} \\
\hiepassert{$~~ \heap_{99} \heq \hone{x}{t_1} ~\hsep~ \heap_1 ~\}$}
}
\end{center}
\caption{Demonstrating Symbolic Execution}
\label{fig:heap-unchanged}
\end{figure}

\noindent
We demonstrate the usefulness (and partly the correctness)
of Proposition~\ref{thm:sp} with a simple example. Consider:

\vspace*{0mm}
\begin{center}
$\oldtriple{\heap_{99} \heq \heapg}{\texttt{$*$x += 1; $*$x -= 1;}}{\heap_{99} \heq \heapg}$
\end{center}
\vspace*{0mm}

\noindent
In other words, the heap is unchanged after an increment and then a decrement.
We rewrite the program so that only one heap
operation is performed per program statement; in
Fig.~\ref{fig:heap-unchanged} we show the rewritten program fragment
together with the propagation of the formulas.  (For brevity, we also
perform a simplification step.)  It
is then easy to show that the final formula implies
$\heap_{99} \heq \heapg$, by first establishing that
$\heap_1 \heq \heap_3 \heq \heap_4$ and $v = t_2 = t_1 +1$.  
This example provides a program \emph{summary}
that the heap is the same before and after execution.

\section{The Frame Rule}
\label{sec:framerule}

Recall the classic frame rule (CFR) from Section \ref{sec:intro} where
from $\oldtriple{\phi}{P}{\psi}$ we may infer
$\oldtriple{\phi \wedge \pi}{P}{\psi \wedge \pi}$ with the side
condition that $P$ does not modify any free variable in $\pi$.  In our
current setting where $P$ now may contain heap references, this frame
rule in fact \emph{still} can be used if $\pi$ only contains free heap
variables that are \emph{ghost}.  However, because the global heap
memory might  be changed by $P$, what \emph{cannot} be framed
through with this rule, is the property that a ghost variable $\heap$
is consistent with the global heap memory $\heapg$, i.e.,
$\heap \sqsubseteq \heapg$.  We call such a property the ``heap
reality'' of $\heap$.

In Separation Logic, where heaps are of the main interest, a key step
is that when a program fragment is ``enclosed'' in some heap, then any
formula $\pi$ whose ``footprint'' is separate from this heap can be
framed through.  Recall the SL frame rule (SFR) from
Section \ref{sec:intro} wherein
the premise $\oldtriple{\phi}{P}{\psi}$ ensures that the
implicit heap arising from the formula $\phi$ captures all the heap
accesses, read or write, in the program fragment $P$.  Therefore
$\oldtriple{\phi ~\hsep~ \pi}{P}{\psi ~\hsep~ \pi}$
naturally follows.

In our setting of explicit heaps, the frame rule, suitably translated
into this language, is simply \emph{not valid} without additional
machinery to ensure enclosure. The concept of enclosure is to have an
explicit subheap (or subheaps) to contain the program
heap updates.  These updates are defined to be the cells that the
program \emph{writes to}, or \emph{deallocates}.  This is because the
property $\heap \sqsubseteq \heapg$, where $\heap$ is a ghost
variable, is falsified \emph{just in case} the program has written to
or deallocated some cell in $\heapg$ whose address is also in
$\dom(\heap)$.  Thus, the heap reality of $\heap$ is lost.  Note that
$\mathbf{malloc}$ changes $\heapg$, but it does not affect what already
in $\heapg$.

\begin{definition}[Heap Update]
Given an address value $v$, a \emph{heap update} to location $v$ is
defined as a statement that either \emph{writes to}
or \emph{deallocates} the location $v$.  $\Box$
\end{definition}

\noindent
Before formalizing our notion of ``enclosure'', however, we first need
a concept of heap ``evolution''.  Let us use the notation
$\tilde{\heap}$ to denote the union $\bigcup_i \heap_i$ of a
collection of subheaps $\heap_1, \cdots , \heap_n$, $n \geq 2$.  Thus
for example, $\tilde{\heap} \sqsubseteq \heapg$ simply abbreviates
$\heap_1 \sqsubseteq \heapg \wedge \cdots \wedge \heap_n \sqsubseteq \heapg$.

\begin{definition}[Evolution]\label{def:evolve}
Given a valid triple $\oldtriple{\phi}{P}{\psi}$, we say that a
collection $\tilde{\heap}$ in $\phi$, where
$\phi \models \tilde{\heap} \sqsubseteq \heapg$,
\emph{evolves} to a collection $\tilde{\heap'}$ in $\psi$,
where $\psi \models \tilde{\heap'} \sqsubseteq \heapg$,
if for each model ${\cal I}$ of $\phi$, executing $P$ from ${\cal I}$ 
will result in ${\cal I'}$, 
such that for any (address) value $v$,
$v \in (\dom({\cal I}(\heapg)) ~\setminus~ \dom({\cal I}(\tilde{\heap})))$ implies
$v \not\in \dom({\cal I'}(\tilde{\heap'}))$.

We shall use the notation
$\oldtriple{\phi}{P}{\psi} \leadsto \SEvolve(\tilde{\heap}, \tilde{\heap'})$ 
to denote such evolution. 
$\Box$
\end{definition}

\noindent
Intuitively,
$\oldtriple{\phi}{P}{\psi} \leadsto \SEvolve(\tilde{\heap}, \tilde{\heap'})$
means that the largest $\tilde{\heap'}$ can be is $\tilde{\heap}$ plus
any new cells allocated by $P$, and minus any that are freed by $P$.
Note also that because the triple is valid, ${\cal I'}$ will be a
model of $\psi$.  One important usage of the evolution concept is as
follows:
any heap $\heap_i$ such that $\heap_i ~\hsep~ \tilde{\heap}$ and
$\heap_i \sqsubseteq \heapg$ at the point of the precondition $\phi$
(i.e., before $P$ is executed), $\heap_i$ will be separate from
$\tilde{\heap'}$ at the point of the postcondition (i.e., after $P$ is
executed).

Consider the (linked-list) {\tt node} defined in
Section~\ref{sec:heapformulas} (Fig.~\ref{fig:ex:inc}) and the triple
shown below.

\vspace*{3mm}
\begin{center}
\stuff{
$\hiepassert{\{~{list}(\heap_1, x) \hiepand \heap_1 \sqsubseteq \heapg ~\}}$ \\
\> z = malloc(sizeof(struct node)); \\
\> z->next = x; \\
$\hiepassert{\{~{list}(\heap'_1, z) \hiepand  \heap'_1 \sqsubseteq \heapg ~\}}$ 
}
\end{center}

\noindent
We say that $\heap'_1$ is an \emph{evolution}
of $\heap_1$, or  $\SEvolve(\heap_1, \heap'_1)$, notationally. Now
assume that the triple represents only a local proof (i.e., we are also
interested in other parts of $\heapg$).  How should we compose this
local triple to obtain a new triple?
Formally, we have the following:

\vspace*{3mm}
{\centering
$\begin{array}{c}
\infer{\oldtriple{\phi ~\land~ \tilde{\heap}~\hsep~\heap_0
~\land~\heap_0 \sqsubseteq \heapg}{P}{\psi ~\land~ \tilde{\heap'}~\hsep~\heap_0}}
{\oldtriple{\phi}{P}{\psi} \leadsto \SEvolve(\tilde{\heap}, \tilde{\heap'})
}\end{array}$\hfill ({\sc EV})
}
\vspace*{1mm}

\begin{theorem}[Propagation of Separation]
The rule \emph{({\sc EV})} is correct.
$\Box$
\end{theorem}

\begin{myproof}
Let ${\cal I}$ be a model of $\phi$ that is also a model of
$\tilde{\heap}~\hsep~\heap_0 ~\land~\heap_0 \sqsubseteq \heapg$.
Let ${\cal I'}$ be the result of executing $P$ from ${\cal I}$.
For each address $v \in \dom({\cal I'}(\heap_0))$, 
because $\heap_0$ is a ghost variable, i.e., 
its domain is not affected by executing $P$, 
we also have $v \in \dom({\cal I}(\heap_0))$.
It follows that $v \in (\dom({\cal I}(\heapg)) ~\setminus~ \dom({\cal I}(\tilde{\heap})))$.
Directly from the definition of evolution, we deduce
$v \not\in \dom({\cal I'}(\tilde{\heap'}))$ must hold.
As a result, ${\cal I'}$ also satisfies $\tilde{\heap'} ~\hsep~ \heap_0$.
$\Box$
\end{myproof}

\ignore{
\begin{proof}
(Sketch.) 
Let ${\cal I}$ be a model of $\phi$ that is also a model of
$\tilde{\heap}~\hsep~\heap_0 ~\land~\heap_0 \sqsubseteq \heapg$.
Let ${\cal I'}$ be the result of executing $P$ from ${\cal I}$.
For each address $v \in \dom({\cal I'}(\heap_0))$, 
because $\heap_0$ is a ghost variable, i.e., 
its domain is not affected by executing $P$, 
we also have $v \in \dom({\cal I}(\heap_0))$.
It follows that $v \in (\dom({\cal I}(\heapg)) ~\setminus~ \dom({\cal I}(\tilde{\heap})))$.
Directly from the definition of evolution, we deduce
$v \not\in \dom({\cal I'}(\tilde{\heap'}))$ must hold.
As a result, ${\cal I'}$ also satisfies $\tilde{\heap'} ~\hsep~ \heap_0$.
\end{proof}
}

\noindent
We are now ready to describe our notion of enclosure.
We wish to describe, given a program $P$ and 
a heap collection $\tilde{\heap}$ in a precondition 
description $\phi$, that all heap updates
(heap assignments or deallocations) in $P$,
are confined to an evolution of $\tilde{\heap}$.
The following definition, intuitively, is about one
aspect of memory-safety: the heap updates are safe.

\begin{definition}[Enclose]\label{def:enclose}
Suppose we have a valid triple $T = \oldtriple{\phi}{P}{\_}$,
$\tilde{\heap}$ appears in $\phi$, and that
$\phi \models \tilde{\heap} \sqsubseteq \heapg$.
We say $\tilde{\heap}$ \emph{encloses} all heap updates of $P$ if 
for any model ${\cal I}$ of $\phi$ and 
for any execution path of $P$ of the form 
$P_1; s; P_2$ where $s$ is a heap update to a location $v$, 
it follows that there exists $\tilde{\heap'}$ s.t.   
$\oldtriple{\phi}{P_1}{\_} \leadsto \SEvolve(\tilde{\heap}, \tilde{\heap'})$ and  
$v \in \dom({\cal I'}(\tilde{\heap'}))$ hold, where ${\cal I'}$ is the result of 
executing $P_1$ from ${\cal I}$.

We shall use the notation $T \leadsto \SEnclose(\tilde{\heap})$
to denote that $\tilde{\heap}$ encloses all the updates of $P$ wrt. $T$.
$\Box$
\end{definition}

\noindent
We now can introduce our frame rule.  It is in fact all about
``preserving the heap reality''.  Recall that a recursive constraint,
which satisfies the standard side condition and of which the heap
variables are all ghost (and this is a common
situation), \emph{remains true} from precondition to postcondition.
What may no longer hold in the postcondition is the heap reality of
some $\heap_0$. That is, $\heap_0 \sqsubseteq \heapg$ may hold
at the precondition, but no longer so at the postcondition.  In other
words, given local reasoning for a code fragment $P$ and the fact that
$\heap_0 \sqsubseteq \heapg$ holds before executing $P$, how would we
preserve this heap reality, without the need to reconsider the code
fragment $P$?  Our answer is the following Hoare-style rule, our new
frame rule:

\vspace*{3mm}
\hspace{-5mm}
$\begin{array}{c}
\infer{\oldtriple{\phi ~\land~ \tilde{\heap}~\hsep~\heap_0 ~\land~ \heap_0 \sqsubseteq \heapg}{P}{\psi ~\land~ \heap_0 \sqsubseteq \heapg}}{\oldtriple{\phi}{P}{\psi} \leadsto \SEnclose(\tilde{\heap}) 
}
\end{array}$\hfill ({\sc FR})
\vspace*{1mm}

\begin{theorem}[Frame Rule]
The rule \emph{({\sc FR})} is correct.
$\Box$
\end{theorem}

\begin{myproof}
We prove by contradiction. Assume it is not the case, 
meaning that there is  model ${\cal I}$ of $\phi$ that is also a model of
$\tilde{\heap}~\hsep~\heap_0 ~\land~\heap_0 \sqsubseteq \heapg$ and
${\cal I'}$ is the result of executing $P$ from ${\cal I}$, but 
${\cal I'}$ does not satisfy $\heap_0 \sqsubseteq \heapg$.
Thus there must be a cell $\one{v}{\_}$ that 
belongs to ${\cal I'}(\heap_0)$ but not ${\cal I'}(\heapg)$.
Because ${\cal I}(\heap_0) \sqsubseteq {\cal I}(\heapg)$, 
the fragment $P$ must have updated the location $v$. Therefore, there must
be an execution path of $P$ which is of the form $P_1; s; P_2$, where
$s$ is a heap update to the location $v$. 
Let ${\cal \overline{I}}$ be the result of executing $P_1$ from ${\cal I}$.
By the definition of enclosure, assume $\oldtriple{\phi}{P_1}{\_} \leadsto \SEvolve(\tilde{\heap}, \tilde{\heap'})$ 
and $v \in \dom({\cal \overline{I}}(\tilde{\heap'}))$ hold.
By (EV) rule, we have ${\cal \overline{I}}$ satisfies $\tilde{\heap'}~\hsep~\heap_0$.
Since $\heap_0$ is a ghost variable, its domain is not affected by
executing $P_1$, i.e., $v \in \dom({\cal \overline{I}}(\tilde{\heap_0}))$ holds.
This is a contradiction.
$\Box$
\end{myproof}

\noindent
Let us demonstrate the use of the two theorems on a very simple example. 
Consider the triple:

\vspace{1mm}
\begin{center}
\stuff{
$\hiepassert{\{(\hone{x}{\_} ~\hsep~ \heap) \sqsubseteq \heapg\}}$ 
$*$x = 1; 
$\hiepassert{\{(\hone{x}{1} ~\hsep~ \heap) \sqsubseteq \heapg\}}$ 
}
\end{center}
\vspace{2mm}

\noindent
We could follow the symbolic execution rules presented in Section~\ref{sec:symexec} and also
be able to prove this triple. 
But, for the sake of discussion, we consider local reasoning over triple $T$:

\vspace{2mm}
\begin{center}
\colorbox{lightgray}{$\oldtriple{\hone{x}{\_} \sqsubseteq \heapg}{*x = 1;}{\hone{x}{1} \sqsubseteq \heapg}$},
\end{center}
\vspace{2mm}

\noindent
which holds trivially. Also, we can clearly see that both
$T \leadsto \SEvolve(\hone{x}{\_}, \hone{x}{1})$ and
$T \leadsto \SEnclose(\hone{x}{\_})$ hold.  Applying the rule ({\sc
EV}), we deduce that $\hone{x}{1} ~\hsep~ \heap$ holds after executing
the code fragment.  Furthermore, applying the frame rule, rule
({\sc FR}), we deduce that $\heap \sqsubseteq \heapg$ remains true,
i.e., the heap reality of $\heap$ is preserved.  Putting the pieces
together, we can establish the truth of the original triple by making
use of the two theorems.

Recall that we use traditional conjunction, as opposed to separating
conjunction in SL.   We thus emphasize that all the rules presented above 
(CFR, EV and FR in particular) can be used in combination
because in our framework:
$\oldtriple{\phi}{P}{\psi_1}$ and 
$\oldtriple{\phi}{P}{\psi_2}$
imply
$\oldtriple{\phi}{P}{\psi_1 \wedge \psi_2}$.

\ignore{ 

\noindent
We now revisit the {\tt list} example presented earlier in this section,
but in the context that there
exists a tree housed by a heap $\heap_2$ that is separate from
$\heap_1$.  Consider the triple T':

\vspace*{2mm}
\stuff{
$\hiepassert{\{~list(\heap_1, x) \hiepand tree(\heap_2, y) \hiepand \heap_1 \hsep \heap_2 \sqsubseteq \heapg~\}}$ \\
\> z = malloc(sizeof(struct node)); \\
\> *z = x; \\
$\hiepassert{\{~list(\heap_1', z) \hiepand tree(\heap_2, y)
\hiepand \heap'_1 \hsep \heap_2 \sqsubseteq \heapg~\} }$
}
\vspace*{2mm}

\noindent
and let $P$ stand for the
program fragment of interest.  From the fact that 
$T \leadsto \SEvolve(\heap_1, \heap'_1)$ and
$T \leadsto \SEnclose(\hemp)$ hold, and by applying the two rules ({\sc
EV}) and ({\sc FR}), we can easily establish the validity of this triple T'.
}

\floatstyle{boxed}
\restylefloat{figure}

\renewcommand{\arraystretch}{2}

\begin{figure}[tbh!]
\begin{center}
\hspace*{-4mm}
\begin{tabular}{c}
\mystuff{
\begin{array}[t]{c}
\infrule{MALLOC}\\
$\infer{\oldtriple{\phi}{\texttt{x = malloc(1)}}{\psi} \leadsto \SEvolve(\tilde{\heap}, \tilde{\heap'})}
{{\phi \models \tilde{\heap} \sqsubseteq \heapg}
\hspace{10mm} {\psi \models \tilde{\heap'} \sqsubseteq \heapg} \\
{\psi \models \dom( \tilde{\heap'}) \subseteq \dom(\tilde{\heap}) \cup \{ x \}}
}$
\end{array}
}  \\
\mystuff{
\begin{array}[t]{c}
\infrule{FREE}\\
$\infer{\oldtriple{\phi}{\texttt{free(x)}}{\psi} \leadsto \SEvolve(\tilde{\heap}, \tilde{\heap'})}
{{\phi \models \tilde{\heap} \sqsubseteq \heapg}
\hspace{10mm} {\psi \models \tilde{\heap'} \sqsubseteq \heapg} \\
{\psi \models \dom( \tilde{\heap'}) \subseteq \dom(\tilde{\heap}) \setminus \{ x \}}
}$
\end{array}
} \\
\mystuff{
\begin{array}[t]{c}
\infrule{OTHER-STATEMENTS}\\
$\infer{\oldtriple{\phi}{s}{\psi}  \leadsto \SEvolve(\tilde{\heap}, \tilde{\heap'})}
{{\phi \models \tilde{\heap} \sqsubseteq \heapg}
\hspace{10mm} {\psi \models \tilde{\heap'} \sqsubseteq \heapg} \\
{\psi \models \dom( \tilde{\heap'}) \subseteq \dom(\tilde{\heap})}
}$
\end{array}
} \\
\mystuff{
\begin{array}[t]{c}
\infrule{SEQ-COMPOSITION}\\
$\infer{\oldtriple{\phi}{P; Q}{\gamma} \leadsto \SEvolve(\tilde{\heap}, \tilde{\heap''})}
{\oldtriple{\phi}{P}{\psi} \leadsto \SEvolve(\tilde{\heap}, \tilde{\heap'}) \hfill \\
{\oldtriple{\psi}{Q}{\gamma} \leadsto \SEvolve(\tilde{\heap'}, \tilde{\heap''})}
}$
\end{array}
} \\
\mystuff{
\begin{array}[t]{c}
\infrule{CALL}\\
$\infer{\oldtriple{\phi'}{\texttt{call~f()}}{\_}  \leadsto \SEvolve(\tilde{\heap}, \tilde{\heap'})}
{
{[\oldtriple{\phi}{{\tt f()}}{\psi}  \leadsto \SEvolve(\tilde{\heap}, \tilde{\heap'})] \in Specs} 
\hspace{10mm} {\phi' \models \phi}
}$
\end{array}
} \\
\mystuff{
\begin{array}[t]{c}
\infrule{COMPOSITION}\\
$\infer{\oldtriple{\phi}{P}{\psi}  \leadsto
  \SEvolve(\tilde{\heap_1} \cup \tilde{\heap_2}, \tilde{\heap'_1} \cup \tilde{\heap_2})}
{
  \oldtriple{\phi}{P}{\psi} \leadsto \SEvolve(\tilde{\heap_1}, \tilde{\heap'_1}) \hfill \\
  \oldtriple{\phi}{P}{\psi} \leadsto \SEnclose(\tilde{\heap}) \hspace{10mm}
  {\phi \models \tilde{\heap} ~\hsep~ \tilde{\heap_2} \wedge \tilde{\heap_2} \sqsubseteq \heapg}
}$
\end{array}
}
\end{tabular}
\end{center}
\vspace{-2mm}
\caption{Hoare-style Rules for Evolution. {\scriptsize OTHER-STATEMENTS} applies
to \texttt{s} not of the kind covered by the rules above.}
\label{fig:evolve}
\end{figure}

\begin{figure}[tbh!]
\begin{center}
\hspace*{-4mm}
\begin{tabular}{c}
\mystuff{
\begin{array}[t]{c}
\infrule{HEAP-ASSIGN}\\
$\infer{\oldtriple{\phi}{\texttt{$*$x = y}}{\_} \leadsto \SEnclose(\tilde{\heap})}
{{\phi \models \tilde{\heap} \sqsubseteq \heapg}
\hspace{10mm} {x \in \dom(\tilde{\heap})}
}$
\end{array}
}
\\
\mystuff{
\begin{array}[t]{c}
\infrule{FREE}\\
$\infer{\oldtriple{\phi}{\texttt{free(x)}}{\_} \leadsto \SEnclose(\tilde{\heap})}
{{\phi \models \tilde{\heap} \sqsubseteq \heapg}
\hspace{10mm} {x \in \dom(\tilde{\heap})}
}$
\end{array}
}
\\
\mystuff{
\begin{array}[t]{c}
\infrule{OTHER-STATEMENTS}\\
$\infer{\oldtriple{\phi}{s}{\_}  \leadsto  \SEnclose(\tilde{\heap})}
{{\phi \models \tilde{\heap} \sqsubseteq \heapg}
}$
\end{array}
}
\\
\mystuff{
\begin{array}[t]{c}
\infrule{SEQ-COMPOSITION}\\
$\infer{\oldtriple{\phi}{P;Q}{\gamma}  \leadsto \SEnclose(\tilde{\heap})}
{
\oldtriple{\phi}{P}{\psi} \leadsto \SEvolve(\tilde{\heap}, \tilde{\heap'}) \hfill \\
\oldtriple{\phi}{P}{\psi} \leadsto \SEnclose(\tilde{\heap}) \hspace{3mm}
\oldtriple{\psi}{Q}{\gamma} \leadsto \SEnclose(\tilde{\heap'}) \hfill
}$
\end{array}
}
\\
\mystuff{
\begin{array}[t]{c}
\infrule{CALL}\\
$\infer{\oldtriple{\phi'}{\texttt{call~f()}}{\_}  \leadsto \SEnclose(\tilde{\heap})}
{{[\oldtriple{\phi}{\texttt{f()}}{\psi} \leadsto \SEnclose(\tilde{\heap})] \in Specs} \\
{\phi' \models \phi \wedge \tilde{\heap} \sqsubseteq \heapg} \hfill
}$
\end{array}
}
\\
\mystuff{
\begin{array}[t]{c}
\infrule{COMPOSITION}\\
$\infer{\oldtriple{\phi}{P}{\psi}  \leadsto \SEnclose(\tilde{\heap} \cup \tilde{\heap'})}
{
\oldtriple{\phi}{P}{\psi} \leadsto \SEnclose(\tilde{\heap}) \hspace{10mm}
{\phi \models \tilde{\heap'} \sqsubseteq \heapg}
}$
\end{array}
}
\\
\end{tabular}
\end{center}
\vspace{-2mm}
\caption{Hoare-style Rules for Enclosure. {\scriptsize OTHER-STATEMENTS} applies
to \texttt{s} not of the kind covered by the rules above.}
\label{fig:enclose}
\end{figure}

\vspace{2mm}
\noindent
{\bf Our frame rules vs. SL's frame rule:}
We now elaborate the connection of our two rules ({\sc EV}) 
and ({\sc FR}) with the traditional frame rule in Separation Logic (SL).
First, why do we have two rules while SL has one, as introduced in the
beginning of this section?  The reason is that
SL, succinctly, captures \emph{two} important properties: that

\vspace*{2mm}
\begin{itemize}
\setlength\itemsep{0mm}
\item $\pi$ can be added to precondition $\phi$
and it \emph{remains true} in the postcondition;
\item $\pi$ \emph{retains its separateness}, from precondition $\phi$ to 
postcondition $\psi$.
\end{itemize}
\vspace*{2mm}

\noindent
The second property is important for \emph{successive} uses of the
frame rules.  Our rule ({\sc FR}) above only provides for the first
property.  We accommodate the second property with the other rule
({\sc EV}), i.e., the ``propagation of separation'' rule.

The two concepts of evolution and enclosure in fact
exist in SL, \emph{implicitly}.  Given the triple $T
= \oldtriple{\phi}{P}{\psi}$, assume that $\heap$ is the heap housing
the precondition $\phi$ and $\heap'$ is the heap housing the
postcondition $\psi$.  In SL, the frame rule also requires that
$T \leadsto \SEvolve(\heap, \heap')$ and that $T \leadsto \SEnclose(\heap)$.
In short, this means that whenever the traditional frame rule in SL\footnote{We
assume an SL fragment without magic wands.} is applicable,
our frame rules are also applicable without any additional complexity.

However, in general our assertion language allows for multiple
subheaps, which entails more expressive power, but at the cost that we
no longer can resort to the abovementioned default.  For this paper, we require
the specifications to also nominate the subheaps participating in the
evolution and/or enclosure relations, stated under
the keyword {\bf frame}, following the typical {\bf requires} and {\bf
ensures} keywords.  We demonstrate this in Section~\ref{sec:drive}
with our driving example.

\vspace{2mm}
\noindent
{\bf Proving the Evolution and Enclosure relations.}
The next question of interest is how the evolution and enclosure
relations are practically checked.  For evolution, we use the rules in
\mbox{Fig.~\ref{fig:evolve}}. In the rule \rulen{CALL},

\vspace{2mm}
\begin{center}
\colorbox{lightgray}{${[\oldtriple{\phi}{{\tt
f()}}{\psi} \leadsto \SEvolve(\tilde{\heap}, \tilde{\heap'})] \in Specs}$}
\end{center}
\vspace{2mm}

\noindent
means that we have nominated
$\SEvolve(\tilde{\heap}, \tilde{\heap'})$ the specifications of
function {\tt f}. Similarly for enclosure relation,
which can be effectively checked
using the rules presented in \mbox{Fig.~\ref{fig:enclose}}.
Checking evolution and enclosure relations is also performed modularly. 
Specifically, at call sites, we make use of the rule \rulen{CALL} and then
achieve compositional reasoning with the rule \rulen{COMPOSITION}.

We finally conclude this section with two Lemmas about the correctness
of the rules presented in Figures~\ref{fig:evolve} and~\ref{fig:enclose}.
The proofs of the two lemmas follow similar (but more tedious) steps as 
in proving our two main theorems. For brevity, we omit the details.

\vspace*{2mm}
\begin{lemma}[Evolution]
Given a valid triple $T = \oldtriple{\phi}{P}{\psi}$ where
$\phi \models \tilde{\heap} \sqsubseteq \heapg$  and
$\psi \models \tilde{\heap'} \sqsubseteq \heapg$,
$T \leadsto \SEvolve(\tilde{\heap}, \tilde{\heap'})$ holds if it
follows from the rules in \mbox{Fig.~\ref{fig:evolve}}.  $\Box$
\end{lemma}

\vspace*{2mm}
\begin{lemma}[Enclose]
Given a valid triple $T = \oldtriple{\phi}{P}{\_}$ where
$\phi \models \tilde{\heap} \sqsubseteq \heapg$,
$T \leadsto \SEnclose(\tilde{\heap})$
holds if it follows from the rules in \mbox{Fig.~\ref{fig:enclose}}.
$\Box$
\end{lemma}



\section{Automatic Proof of the Challenge Example}
\label{sec:drive}

\begin{figure}[hb]
\mystuff{
mgraph($h, x, t^{in}$) \clpif \\
\> $h \heq \hemp, x = \Null$. \\
mgraph($h, x, t^{in}$) \clpif \\
\> $\hone{x}{({\color{blue} 1}, \_, \_)} \sqsubseteq t^{in}$, $h \heq \hemp$. \\
mgraph($h, x, t^{in}$) \clpif \\
\> $h_x \heq \hone{x}{({\color{blue} 1}, l, r)}$, $t_l \heq h_x ~\hsep~ t^{in}$, \\
\>  mgraph($h_{l}, l, t_l$), $t_r \heq  h_{l} ~\hsep~ t_l$, \\
\>  mgraph($h_{r}, r, t_r$), $h \heq h_x ~\hsep~ h_{l} ~\hsep~ h_{r}$, $h ~\hsep~ t^{in}$.  \\

\ \\

pmg($h, x, t^{in}, t^{out}$) \clpif \\
\> $h \heq \hemp$, $x = \Null$, $t^{out} \heq t^{in}$. \\
pmg($h, x, t^{in}, t^{out}$) \clpif \\
\> $\hone{x}{({\color{blue} 1}, \_, \_)} \sqsubseteq t^{in}$, $h \heq \hemp$, 
$t^{out} \heq t^{in}$. \\
pmg($h, x, t^{in}, t^{out}$) \clpif  \\
\> $h_x \heq \hone{x}{({\color{blue} 1}, l, r)}$,
$t_l \heq h_x ~\hsep~ t^{in}$, \\
\>  mgraph($h_{l}, l, t_l$), $t_r \heq h_{l} ~\hsep~ t_l$, mgraph($h_{r}, r, t_r$), \\
\> $t_{out} \heq h_{r} ~\hsep~ t_r$, $h \heq h_x ~\hsep~ h_{l} ~\hsep~ h_{r}$, $h ~\hsep~ t^{in}$.\\
pmg($h, x, t^{in}, t^{out}$) \clpif \\
\> $h_x \heq \hone{x}{({\color{blue} 0}, l, r)}$, 
$t_l^{in} \heq \hone{x}{({\color{blue} 1}, l, r)} ~\hsep~ t^{in}$, \\
\> pmg($h_l, l, t_{l}^{in}, t_r^{in}$), pmg($h_{r}, r, t_r^{in}, t^{out}$), \\
\> $h \heq h_x ~\hsep~ h_{l} ~\hsep~ h_{r}$,  $h ~\hsep~ t^{in}$.
\ \\
}
\vspace{-3mm}
\caption{Definitions of \texttt{mgraph} and \texttt{pmg}}\label{fig:def:mark}
\end{figure}



Recall the graph marking algorithm in Fig.~\ref{fig:markgraph}, which
exhibits precisely four scenarios.  (1) The function terminates upon
seeing a $\Null$ pointer.  The function also terminates upon
encountering a marked node. For this there are two possibilities: (2)
the current node has been encountered before (in the history); or (3)
the subgraph rooted at the current node had already been fully marked
(modulo the history).  Finally, when encountering an unmarked node
(4), the function first marks the node, then invokes two recursive
calls to deal with the left and right subgraphs. This last scenario
poses a technical challenge, concerning separation of the two
recursive calls, so that a frame rule can be used to protect the
effects of the first call from the that of the second.  In actual
fact, the second call can refer to a portion of the heap modified by
the first call.  The important point however is the second call does
not \emph{write} to this subheap.


The four rules in our definition of ``partially-marked-graph''
\texttt{pmg}($h,x,t^{in}, t^{out}$) correspond to the four scenarios
identified above.  We address the technical challenges by having: (a)
$h$ encloses the \emph{write} footprint of the code while precisely
excludes the nodes that had been visited in the history; (b) $t^{in}$
captures the history, i.e. nodes visited starting from the root node
to the current node $x$; and importantly, (c) $t^{out}$ captures the
output history, which would be the set of visited nodes right after
the function {\tt mark} finishes processing the subgraph rooted
at $x$.  The use of $t^{out}$ resembles a form of ``continuation
passing''.

The importance of $t^{out}$ can be understood by investigating the
$4^{th}$ scenario identified above. Encountering the node $x$ that is
unmarked, the function first marks it before recursively processing
the left subgraph and then the right subgraph. What then should be
used as the histories for these recursive calls? The history used for
the first call can be easily constructed by conjoining the history of
the call to $x$ with the updated node $x$ (the mark field has been
set). However, the actual history used for the second call very much
depends on the shape of the original graph. We choose to construct
$t^{out}$ recursively, thus the output history of the first recursive
call can be conveniently used as the input history for the second
call. The $4^{th}$ rule in the definition of \texttt{pmg} closely
follows these intuitions.

Before proceeding, we contrast here our use of the predicate
\texttt{pmg} with the way predicates are used in SL.  In SL, a
predicate describes (a part of) the current heap; in \texttt{pmg}, we
simultaneously describe \emph{three heaps} corresponding to different
stages of computation.

\pppreset
\begin{figure*}[tbh]
\begin{center}
\stuff{
{\bf requires:} \> \> \> \>  $ \texttt{pmg}(\heap, x, t^{in}, t^{out}) \hiepand \heap \sqsubseteq \heapg \hiepand t^{in} \sqsubseteq \heapg$ \\
{\bf ensures:} \> \> \> \>  $\texttt{mgraph}(\heap', x, t^{in}) \hiepand \heap' \sqsubseteq \heapg \hiepand t^{out} \heq t^{in} ~\hsep~\heap' \hiepand$ \\
{\bf frame:}  \> \> \> \> $\Enclose(\heap) \hiepand  \Evolve(\heap, \heap')$ \\ \ \\
void mark(struct node $*$x) \{ \\
$\hiepassert{ \{~pmg(\heap, x, t^{in}, t^{out}) \hiepand \heap \sqsubseteq \heapg \hiepand t^{in} \sqsubseteq \heapg ~\}}$ \\
\pppred \> assume(x \&\& x->m != 1); struct node *l = x->left, *r = x->right; \\
$\hiepassert{\{~\heap_x \heq \hone{x}{(0, l, r)} \hiepand t_{l}^{in} \heq \hone{x}{(1, l, r)}  ~\hsep~ t^{in}  \hiepand pmg(\heap_{l}, l, t_{l}^{in}, t_r^{in}) \hiepand pmg(\heap_{r}, r, t_r^{in}, t^{out}) \hiepand}$ \\
$\hiepassert{~~ \heap \heq \heap_x ~\hsep~ \heap_{l} ~\hsep~ \heap_{r} \hiepand \heap ~\hsep~ t^{in} \hiepand \heap \sqsubseteq \heapg \hiepand t^{in} \sqsubseteq \heapg ~\}}$ \\
\pppred \> x->m = 1; \\
$\hiepassert{\{~ pmg(\heap_{l}, l, t_{l}^{in}, t_r^{in}) \hiepand t_{l}^{in} \heq \hone{x}{(1, l, r)} * t^{in} \hiepand \heap_{l} * \heap_{r} *  t_{l}^{in} \sqsubseteq \heapg \hiepand  pmg(\heap_{r}, r, t_r^{in}, t^{out})~\}} $ \\ 
\pppred \> mark(l); \\
$\hiepassert{ \{~mgraph(\heap'_{l}, l, t_{l}^{in}) \hiepand  \heap'_{l} \sqsubseteq \heapg \hiepand  t_r^{in} \heq t_{l}^{in} ~\hsep~ \heap'_{l} \hiepand}$ \hspace{25mm} // postcondition \\
$\hiepassert{~~ t_{l}^{in} \heq \hone{x}{(1, l, r)} * t^{in} \hiepand pmg(\heap_{r}, r, t_r^{in}, t^{out}) \hiepand} $ \hspace{27mm} // {\sc (CFR)} \\
$\hiepassert{~~ \colorbox{lightgray}{$\heap'_{l} ~\hsep$}  \heap_{r} * t_{l}^{in} \hiepand }$ \hspace{72mm} // {\sc (EV)} \\  
$\hiepassert{~~\colorbox{lightgray}{$\heap_{r} ~\hsep~ t_{l}^{in} \sqsubseteq \heapg$}~\}} $ \hspace{71mm} // {\sc (FR)} \\ 
\pppred \> mark(r); \\
$\hiepassert{\{~ mgraph(\heap'_{l}, l, t_{l}^{in}) \hiepand  t_r^{in} \heq t_{l}^{in} ~\hsep~ \heap'_{l} \hiepand  t_{l}^{in} \heq \hone{x}{(1, l, r)} * t^{in} \hiepand }$ 
\hspace{3mm} // {\sc (CFR)} \\  
$\hiepassert{~~ mgraph(\heap'_{r}, r, t_r^{in}) \hiepand \heap'_{r} \sqsubseteq \heapg \hiepand  t^{out} \heq t_r^{in} ~\hsep~ \heap'_{r} \hiepand }$  \hspace{24mm} // postcondition \\
$\hiepassert{~~ \colorbox{lightgray}{$\heap'_{r} ~\hsep~$} \heap'_{l} ~\hsep~ t_{l}^{in} \hiepand  }$   \hspace{68mm} // {\sc (EV)} \\ 
$\hiepassert{~~ \colorbox{lightgray}{$\heap'_{l} \sqsubseteq \heapg \hiepand t_{l}^{in} \sqsubseteq \heapg $}~\}} $  \hspace{64mm} //  {\sc (FR)} \\
\}
}
\vspace{-1mm}
\caption{Mark Graph Example}
\label{fig:mark:proof}
\end{center}
\end{figure*}

In Fig.~\ref{fig:mark:proof} we show the specification of the function
\texttt{mark} and the proof for the most interesting case: $x$ is not
$\Null$ and its \texttt{m} field has not been marked.  In the
precondition, ${\heap}$, the first component of the definition of
\texttt{pmg}, appropriately encloses the \emph{write} footprints of
the function. It is thus easy to derive, \Enclose{}$(\heap)$.  Proving
that \Evolve{}$(\heap, \heap')$ is also standard, thus we will not
elaborate on this. Instead, let us focus the discussion on how the
frame rules are used.

The assertion after step {\color{red} \texttt{1}} is obtained
by unfolding the definition of \texttt{pmg} using the
fourth rule and instantiating the values of $l$ and $r$.  Note that
this unfolding is triggered since the footprint of $x$ is touched.
(Using the other rules will lead to a conflict with the
guard \texttt{assume(x \&\& x->m != 1)}.)  At the recursive
call \texttt{mark(l)} (point {\color{red} \texttt{3}}), we need to
prove that the assertion after program point {\color{red} \texttt{2}}
implies the precondition of the function {\tt mark}.
In this context, the precondition is:
\colorbox{lightgray}{$\texttt{pmg}(\heap^{l}, l, {t^{in}}^{l}, {t^{out}}^{l}) \hiepand 
\heap^{l} \sqsubseteq \heapg \hiepand {t^{in}}^{l} \sqsubseteq \heapg$}.
Such a proof can be achieved simply by matching $\heap^{l}$ with $\heap_{l}$,
 ${t^{in}}^{l}$ with $t_{l}^{in}$, and ${t^{out}}^{l}$ with $t_{r}^{in}$.

The assertion after this call (step {\color{red} \texttt{3}})
is then obtained by application of framing.
First we use the specification to
replace the first occurrence of \texttt{pmg} by \texttt{mgraph}.
What we would like to focus on here is the shaded
heap formula. First, applying rule \textsc{(FR)}, we frame 
\colorbox{lightgray}{$\heap_{r} ~\hsep~ t_{l}^{in} \sqsubseteq \heapg$}
through the step {\color{red} \texttt{3}} because the heaps $\heap_{r}$ and $t_{l}^{in}$
lie outside the updates of the recursive call
\texttt{mark(l)}; note that $\heap_{l} \sqsubseteq \heapg$, however, 
no longer holds and is removed. Second, $\heap$ evolves
into $\heap'$, so a heap's separation from
$\heap$ before the step was propagated into its
separation from $\heap'$ after the step, shown as the 
application of rule \textsc{(EV)}.

This explanation is easily adapted for the call at program point
{\color{red} \texttt{4}}.  Finally, the postcondition is proved by unfolding
\texttt{mgraph}$(\heap', x, t^{in})$ using the third rule, 
followed by appropriate variable matching.

\ignore{ 123

One might prefer another definition of a (fully) marked graph, for example, 
by defining that graph as a collection of edges,
and that all its connecting vertices are marked.
This definition does not reflect the above mentioned assumption
of the programmer, thus it is not appropriate to immediately 
act as the postcondition for the recursive function
{\tt mark}. Instead, we could connect \texttt{mgraph},
where $t$ is instantiated to an empty set, to the new definition.
Ultimately, it is about connecting the specifications of the code
with some ``declarative'' specifications that are independent of any code.
We stress here that this final step is a theorem-proving issue, which
arises frequently in practice, e.g., when verifying sorting algorithms.
Such an issue is beyond the scope of this paper.
Finally, we recall the known fact that the SL frame rule does not 
cover the case when heap cells modified by a code fragment (e.g. a recursive call) 
are \emph{traversed but not modified} by a subsequent fragment.   
This scenario is common in algorithms working on graphs or DAGs;
our frame rule accommodates such a scenario. 

\vspace*{3mm}
NEW STUFF: \\

First to admit similarity of predicate \texttt{mark-successor-closed} used
as a loop invariant and \texttt{markedgraph} in the postcondition.
The attack is that \texttt{mark-successor-closed} is ``contrived''
by the code (ie. traversal order).

First DEFEND by saying that these precisely describe the sets that are
actually used by the two recursive calls.  Then draw an analogy:

\vspace*{3mm}
\stuff{
i = 1; x = 0; y = 1; \\
while (++i <= n) \{ \\
\> z = y; y += x; x = z \\
\}
}
\vspace*{3mm}

\noindent
We wish to prove that for $n \geq 2$, that $y = fib(n)$ eventuates,
where $fib()$ is defined in the traditional recursive way.
A ``contrived '' loop invariant, ie. one constructed by mimicking the code, would be:

\vspace*{3mm}
$\begin{array}{l}
inv(i, x, y) \leftarrow i \leq 1, y = i \\
inv(i, x, y) \leftarrow i \geq 2, y = x' + y', x = y', inv(i-1, x', y')
\end{array}$
\vspace*{3mm}

\noindent
Using this invariant, we would then have to prove that

\vspace*{3mm}
$inv(n, x, y), n \geq 2 \models y = fib(n)$.
\vspace*{3mm}

\noindent
It is easy to see now that this proof, involving two recursive
definitions $inv()$ and $fib*()$, is extremely challenging in general.
In other words, if there is no connection between the predicate
describing the loop invariant and that of the postcondition, there is
little hope for automation.
} 

In our graph marking example, our ``invariant'' precondition involves
the predicate \texttt{pmg} while the final postcondition involves the
predicate \texttt{mgraph}.  The fact that \texttt{pmg} resembles the
code is coincidental but unsurprising, since it needs to describe the
subheaps relevant to the two recursive calls.  One might argue that
the top-level specification \texttt{mgraph} is contrived so as to be
similar to \texttt{pmg}.  One could notice that the former definition
is ``left-askew'', as the ``history'' used for the right subgraph is
computed by conjoining the footprint and the history of the left
subgraph If instead we had used a ``right-askew'' definition, the
final entailment may become very hard to prove.  In the end, this
paper is ultimately about automation, and not about how we can hide
implementation details and use highly declarative specifications.

\vspace*{2mm}
\noindent
{\bf Remark:} There is a recently published proof \cite{muller16isc}
that considers a similar graph marking algorithm.  By supporting the
construct of \emph{iterated separating conjunction}
\cite{reynolds02separation}, they managed to ``verify challenging
examples such as graph-marking algorithms that so far were beyond the
scope of automated verifiers based on permission logics [such as SL
  and IDFs]''.  The critical difference is that their method
precondition does not require that the input graph to be ``properly
marked''.  For example, it does not allow a fully marked graph to be a
valid input. More specifically, it requires that the root node must be
unmarked.  Furthermore, the postcondition on its own \emph{does not
  imply} the final graph is completely marked.  The crucial point here
is that the proof in \cite{muller16isc} does not prove the same thing
as we do.  As an aside, that proof is not about local reasoning; it
does not use framing at all.  Indeed, the specification even refers to
addresses \emph{outside} its code footprint.  The dynamic frame of the
method, \emph{and} those of its recursive method calls, are all the
same: it represents the one global graph.  In the end,
\cite{muller16isc} does not contribute to the reasoning of recursive
predicates.

\ignore{
The second published proof \cite{leino10dafny2} is about 
the Schorr-Waite algorithm. 
However, the program considered is completely different: it comprises a \emph{single} 
non-recursive function and so it has just one dynamic frame.
Hence the proof is not concerned about the two technical points
we are so concerned with: that the input graph is ``properly marked'', allowing for
a mark-successor-closed graph, and the intricate frame reasoning 
when dealing with two successive calls.
}

\ignore{ 444
\subsection{Copy Graph}

\begin{figure}
\hspace{-3mm}
\subfigure[]{
\mystuff{
struct node \{ \\
\> int m; \\
\> struct node *fwd; \\
\> struct node *left; \\
\> struct node *right; \\
\};\\ \ \\ \ \\ \ \\ \ \\
struct node *copy(struct node *x) \{ \\
\> if (!x || x->fwd) \\
\> \> return 0; \\
\> y = malloc(sizeof(struct node)); \\
\> y->fwd = 0; \\
\> struct node *l = x->left; \\
\> struct node *r = x->right; \\
\> y->left = copy(l); \\
\> y->right = copy(r); \\
\> x->fwd = y; \\
\> return y; \\
\}
}
\label{fig:copy-graph}
}
\end{figure}
} 

\ignore{ 888
\hspace{-3mm}
\subfigure[]{
\mystuff{
struct node \{ \\
\> int m; \\ 
\> struct node *left; \\
\> struct node *right; \\
\};\\ \ \\ \ \\ \ \\ \ \\
void mark(struct node *x) \{ \\
\> if (!x || x->m == 1) \\
\> \> return; \\
\> struct node *l = x->left; \\
\> struct node *r = x->right; \\
\> x->m = 1; mark(l); mark(r); \\
\}
}
\label{fig:mark}
}
\hspace{-5mm}
\subfigure[]{
}
} 

\section{A Prototype Implementation}
\label{sec:imp}
We implemented a prototype in \clpr{} \cite{jaffar92clpr}, submitted
as supplementary material for this paper.  


Our prototype takes as input a C program with proper function
specifications.  We support only a small subset of the C programming
language. For example, we support recursive functions but not loops,
and expect loops to be manually compiled into tail-recursive
functions\footnote{We manually translate the example in
Fig.~\ref{fig:ex:inc} to a recursive function, used later as a
benchmark program {\sf increment} in Table~\ref{table:results}.}.
(We also disallow nested expressions, and instead use more temporary
variables.)

\newcommand{\inputf}{$\cal{P}$}
\newcommand{\callf}{$\cal{Q}$}
\newcommand*\Let[2]{#1 $\gets$ #2}
\newcommand{\worklist}{\textsf{States}}
\newcommand{\Iff}{\mbox{\textbf{if}}}
\newcommand{\Break}{\mbox{\textbf{break}}}
\newcommand{\Continue}{\mbox{\textbf{continue}}}
\newcommand{\Endiff}{\mbox{\textbf{endif}}}
\newcommand{\Returnn}{\mbox{\textbf{return}}}
\newcommand{\Elsee}{\mbox{\textbf{else}}}
\newcommand{\Foreach}{\mbox{\textbf{foreach}}}
\newcommand{\Whilee}{\mbox{\textbf{while}}}
\newcommand{\Endfor}{\mbox{\textbf{endfor}}}
\newcommand{\Skip}{\mbox{\textbf{skip}}}
\newcommand{\Then}{\mbox{\textbf{then}}}
\newcommand{\Do}{\mbox{\textbf{do}}}
\newcommand{\To}{\mbox{\textbf{to}}}
\newcommand{\newtransition}[3]{#1 \xrightarrow{#3} #2}

\renewcommand{\true}{\textsf{True}}
\renewcommand{\false}{\textsf{False}}

\newcommand{\checkEntail}{\mbox{\textsc{Entail}}}
\newcommand{\applyFrame}{\mbox{\textsc{FrameRule}}}
\newcommand{\execute}{\mbox{\textsc{SymbolicExec}}}

\begin{algorithm}[h]
\caption{Modular Verification of Pointer Programs}
\label{algo:modular}
\textbf{function} \textsc{Verify}\\
\KwIn{Function \inputf{} as a transition system} 
\KwOut{\true{} if successfully verified; otherwise \false}
\Let{$\phi$}{the precondition of $\cal{P}$} \;
\Let{$\loc_0$}{the starting location of \inputf} \;
\Let{\worklist}{\{ $\pair{\loc_0}{\phi}$ \}} \;

\While{(\worklist{} is not empty)}{
\Let{$\pair{\loc}{\pi}$}{pop a state from \worklist} \;\label{algoline:pop}
\eIf{($\loc$ is ending location of \inputf)}
{
\Let{$\psi$}{the postcondition of \inputf} \; \label{algoline:postbegin}
\Iff~(not \checkEntail($\pi$, $\psi$))~ \Returnn ~\false \; \label{algoline:postend}
}{
\For{each transition $\newtransition{\loc}{\loc'}{\textsf{stmt}}$ in \inputf}{
\eIf{\textsf{stmt} is ``call \callf''}{
\Let{$\phi'$}{the precondition of \callf} \;
\Iff~(not \checkEntail($\pi$, $\phi'$))~ \Returnn ~\false \; \label{algoline:checkpre}
\Let{$\psi'$}{the postcondition of \callf} \;
\Let{$\pi'$}{\applyFrame($\pi$,$\phi'$,$\psi'$)} \;
}{
\Let{$\pi'$}{\execute($\pi$,\textsf{stmt})} \;
}
\Let{\worklist}{$\worklist \cup \{ \pair{\loc'}{\pi'} \}$} \;
}

}
}

\Return ~\true \;
\end{algorithm}

Our prototype compiles a C program into a transition system, which
will be fed into the main algorithm, presented in {\bf
Algorithm~\ref{algo:modular}}.

For presentation purpose, in {\bf Algorithm~\ref{algo:modular}}, 
symbolic state only consists of the current program point and a
{\em state formula} that is in our assertion language presented in
Section~\ref{sec:heapformulas}.

We maintain a worklist of symbolic states ({\sf States}) and consider
one symbolic path at a time (the {\tt while} loop and
line~\ref{algoline:pop}). If we reach the function end point, 
we ensure that the state formula implies the postcondition; if the
entailment does not hold, we can immediately terminate the
verification process with failure
(line \ref{algoline:postbegin}-\ref{algoline:postend}). Otherwise, for
each transition emanating from the current program location $\loc$, we
compute the successor states and add them into the worklist.

We elaborate on the computation of successor states. There are three
types of transitions due to different statement types: (a) a function
call; (b) a statement which manipulates only the {\em stack} memory;
and (c) a heap-manipulating statement as identified in
Section~\ref{sec:heapformulas}.

\begin{itemize}[leftmargin=*]
\item \textbf{For a function call:}  We first check that the precondition of
the corresponding function (\callf) is met, otherwise
we immediately terminate with failure (line~\ref{algoline:checkpre}).
We then apply the frame rules in 
Section~\ref{sec:framerule} in combination with the classical
frame rule (CFR) to achieve compositional reasoning, denoted
by the helper function \applyFrame.
\item  \textbf{For a statement that is not a function call:} We just apply
symbolic execution, denoted by calling the helper function \execute.
Note that for (b), standard symbolic execution is well-understood;
for (c) we used the rules presented in Proposition~\ref{prop:exe}.
\end{itemize}

\ignore{
\vspace{2mm}
\noindent
{\bf For a function call:}  We first check that the precondition of
the corresponding function (\callf) is met, otherwise
we immediately terminate with failure (line~\ref{algoline:checkpre}).
We then apply the frame rules in 
Section~\ref{sec:framerule} in combination with the classical
frame rule (CFR) to achieve compositional reasoning, denoted
by the helper function \applyFrame.

\vspace{2mm}
\noindent
{\bf For a statement that is not a function call:} We just apply
symbolic execution, denoted by calling the helper function \execute.
Note that for (b), standard symbolic execution is well-understood;
for (c) we used the rules presented in Proposition~\ref{prop:exe}.
}

\noindent
For presentation purpose, we omit the details on the handling of
enclosure and evolution relations from our high-level algorithm.  In
fact, the rules in \mbox{Fig.~\ref{fig:evolve}}
and \mbox{Fig.~\ref{fig:enclose}} are incorporated into our
implementation to work in tandem with our symbolic execution rules and
frame rules.  For example, to prove
$\SEvolve(\tilde{\heap}, \tilde{\heap'})$ for a symbolic path, at any
point in the path we would track the largest possible subheap
$\overline{\heap}$ such that
$\SEvolve(\tilde{\heap}, \overline{\heap})$. In the end, the remaining
obligation is to prove that
$\tilde{\heap'} \sqsubseteq  \overline{\heap}$.  For this reason, our
implementation does not suffer from any degree of non-determinism when
dealing with the \rulen{SEQ-COMPOSITION} rules in
Figures~\ref{fig:evolve} and~\ref{fig:enclose}.


%

We now comment on the proofs of {\em entailments} between recursive
definitions at call sites and at the end of a function, i.e., the
helper function \checkEntail.  To demonstrate full automation, our
prototype adapted an entailment check procedure from
\cite{nguyen10shape3,qiu13dryad}.  There they use a general strategy
of unfolding a predicate in both the premise and conclusion until the
entailment becomes obvious; \cite{chu15pldi} describes this strategy
as ``unfold-and-match'' (U+M) and we will follow this terminology.
In particular:

\begin{itemize}[leftmargin=*]
\item
We unfold a recursive constraint on a pointer $x$ when its
    ``footprint'' (e.g., \texttt{x->next}) is touched by the
    code \cite{qiu13dryad}. This step is performed during symbolic
    execution.

\item
At a call site or the end of a function, we deal with obligations of
    the form ${\cal L} \models {\cal R}$, performing a sequence of
    left unfolds (unfolding ${\cal L}$) and/or right unfolds
    (unfolding ${\cal R}$) until the proof obligation is simple enough
    such that a ``proof by matching'' is successful. At this point,
    recursive predicates are treated as \emph{uninterpreted}.  After
    dealing with with the heap constraints, an SMT solver --
    Z3 \cite{z3} -- can be employed to discharge the
    obligation.
\end{itemize}

\noindent
In summary, our entailment check procedure directly deals with
user-defined recursive predicates, yet does {\em not} employ any
user-defined lemmas or axioms.  Neither does it involve newer
technology such as automatic induction \cite{chu15pldi}.  The
important point here is that our entailment check procedure
is obtained not from a \emph{custom} theorem-proving method,
but rather from established methods.

Further, it should also be noted that there are example programs
manipulating linked-lists, whose generated VCs would
require \emph{induction} to prove (see \cite{chu15pldi}). Thus to
build a comprehensive verifier, we should not be content with just
``unfold-and-match''. Instead, we should incorporate other advances in
the area of theorem proving
(e.g. \cite{wies13cav,piskac14cav,chu15pldi}) into the entailment
check procedure \checkEntail.

\renewcommand{\arraystretch}{1.2}
\begin{table}[tb]
\caption{Benchmarking Our Prototype Implementation. \mbox{\texttt{\# VCs}} denotes the number of entailment checks; \mbox{\texttt{\# Z3 calls}} denotes the number of calls to Z3. }

\begin{center}
\begin{tabular}{|l|l|r|r|r|}
\hline
\textsf{Group} & \textsf{Program}  & \textsf{Time (s)} & \texttt{\# VCs} & \texttt{\# Z3 calls} \\
\hline
& {\sf ex1} (Fig.~\ref{fig:heap-unchanged}) & 0.2 & 1 & 11  \\
\textsf{simple} & *{\sf buggy-ex1} & 0.2 & 1 & 11 \\
& {\sf other examples} & 0.3 & 4 & (total) 11 \\
\hline
& {\sf append} & 0.9 & 3 & 86  \\
& {\sf copy} & 8.4 & 3 &  66 \\
& {\sf filter} & 2.0 & 4 & 74 \\
& {\sf increment} & 0.7 & 3 & 58 \\
& {\sf insert} & 0.4 & 1 & 19 \\
\textsf{sll} & {\sf insert-end} & 2.1 & 3 & 74 \\
& {\sf length} & 0.1 & 3 & 23 \\
& *{\sf buggy-length} & 0.2 & 3 & 32  \\
& {\sf remove} & 0.1 & 2 &  15 \\
& {\sf traverse} & 0.1 & 1 &  10 \\
& {\sf zero} & 0.9 & 3 & 60  \\
\hline
 & {\sf bst-search} & 1.0 & 6 & 88 \\
\textsf{tree} & \colorbox{lightgray}{\sf isocopy}& 12.9 & 4 & 75 \\
 & *{\sf buggy-isocopy} & 0.1 & 0 & 23 \\
 & {\sf traverse} & 0.5 & 4 & 38 \\
\hline
 & {\sf markgraph} & 6.8 & 6 & 174 \\
\textsf{graph} & *{\sf buggy-mark1} & 33.7 & 4 & 303 \\
 & *{\sf buggy-mark2} & 11.9 & 4 & 161 \\
\hline
\end{tabular}
\end{center}
\label{table:results}
\end{table}

\vspace{2mm}
\noindent
{\bf Experimental Results.} To demonstrate the applicability of our
framework, other than our breakthrough example and examples presented
throughout this paper, we have also selected a number of example
programs from the \textsc{Grasshopper} system \cite{wies13cav}.  As
sanity checks, we also introduce a number of buggy variants (prefixed
by \texttt{*buggy-}) which, as expected, our prototype will fail to
verify. We used a 2.3 GHz machine running Ubuntu~14.04.3~LTS, with 4GB
memory.  Results appear in Table~\ref{table:results}.

Our benchmarks are in four categories:

\begin{itemize}
\item
{\em heap manipulations} (\textsf{simple}).  The properties
to be proved do not involve recursive constraints. 

\item 
{\em singly-linked lists} (\textsf{sll} for short).  The properties
(collectively) involve reasoning about the shape, data, and size of a
list.

\item
{\em trees}. Programs here traverse a binary and binary search tree.
We also have a distinguished example {\sf isocopy}
which has not been verified before in as general a manner.

\item 
The last group is about our driving example: graph marking, and two
buggy variants.  The purpose here is simply to present some
performance metrics.

\end{itemize}

\ignore{

\begin{figure}[tbh]
\begin{center}
\subfigure[]{
\mystuff{
struct node \{ \\
\> int data; \\
\> struct node *left, *right; \\
\}; \\ 
struct node *copytree(struct node *x) \{ \\
 \> if (!x) return 0; \\
\> y = malloc(sizeof(struct node)); \\
\> y->data = x->data; \\ 
\> y->left = copytree(x->left); \\
\> y->right = copytree(x->right); \\
\> return y; \\
\}
}
\label{fig:copy}
} \\
\hspace{-2mm}
\subfigure[]{
\mystuff{
tree($h, x$) \clpif  ~$h \heq \hemp, x = \Null$. \\
tree($h, x$) \clpif  ~$h \heq (x \mapsto (\_, l, r)) * h_1 * h_2$, \\
\> tree($h_1, l$), tree($h_2, r$).
\ \\ \ \\
isocopy($h_x, x, h_y, y$) \clpif \\
\> $h_x \heq \hemp, x = \Null, h_y \heq \hemp, y = \Null$. \\
isocopy($h_x, x, h_y, y$) \clpif~ $h_x * h_y$, \\
\> $h_x \heq (x \mapsto (d, l_x, r_x)) * h_{xl} * h_{xr}$, \\
\> $h_y \heq (y \mapsto (d, l_y, r_y)) * h_{yl} * h_{yr}$, \\
\> isocopy($h_{xl}, l_x, h_{yl}, l_y$), \\
\> isocopy($h_{xr}, r_x, h_{yr}, r_y$).
}
\label{fig:copyspec}
}
\end{center}
\caption{The \texttt{copytree} example (a); Definitions of \texttt{tree} and \texttt{isocopy} (b)}
\vspace{-3mm}
\end{figure}

}

\vspace{2mm}
\noindent
{\bf Proving isomorphic trees.}  Consider \texttt{isocopy}, which is about the classic problem of
copying a tree.  This program has been previously used
by \cite{berdine05symexec} to demonstrate symbolic execution with
Separation Logic (SL). However, \cite{berdine05symexec} simply proves
that the new tree is separate from the original one; here we prove a
more challenging property, that the copy, also a tree,
is \emph{isomorphic} to the original tree. Specifying such property is
easy using our framework of explicit heaps, because we can simultaneously
describe different heaps corresponding to different stages of
computation.

\ignore{
In
Fig.~\ref{fig:copyspec}, we define a standard binary tree
and \texttt{isocopy}($h_x, x, h_y, y$) stating that the tree in $h_x$
and rooted at $x$ has a separate and isomorphic copy, housed in $h_y$
and rooted at $y$.
Specifying \emph{isomorphism}, which requires recursively relating two subheaps and thus 
is  a challenge for traditional SL, in which each predicate is associated with only a 
single and implicit subheap.

Before proceeding, we note that there have been works, e.g. \cite{ohearn2012natosec},
on specifying isomorphism. They use a separate mathematical formulation 
to define a mathematical tree $\tau$.
Then, such $\tau$ can be used in an assertion such as
$tree(a, \tau) ~\hsep~ tree(b, \tau)$ where
predicate $tree(a, \tau)$ states that $a$
is the root of a tree, which resides in the heap, whose mathematical structure is $\tau$.
Our approach, in contrast, does not require a separate mathematical formulation
for defining $\tau$.  The important consequence of this is that
our logic is \emph{self-contained} and this in turn means that our
assertions can then be effectively propagated through the code in the verification process.

In order to concretize this, consider a follow-up example, below.
Given that the two trees rooted at $a$ and $b$ are isomorphic, 
suppose we change the data at both cells $a$ and $b$ to the same value and
finally, we want to prove that $a$ and $b$ are still isomorphic trees.
Specifying and proving such triple is easy using our framework of explicit heaps,
as follows:


\begin{center}
\stuff{
$\hiepassert{\{~isocopy(\heap_a, a, \heap_b, b) \hiepand \heap_a \sqsubseteq \heapg \hiepand  \heap_b \sqsubseteq \heapg~\}}$ \\
\> assume(a); d = rand(); \\
\> a->data = d; b->data = d;\\
$\hiepassert{\{~isocopy(\heap'_a, a, \heap'_b, b) \hiepand \heap'_a \sqsubseteq \heapg \hiepand  \heap'_b \sqsubseteq \heapg~\}}$
}
\end{center}

\noindent
On the contrary, if a mathematical tree is used, then clearly
we would need two of them: $\tau$ for the precondition and
$\tau'$ for the postcondition.
We would then need to connect the code motion
and the mathematical definitions of these two trees
in order to prove the postcondition. (Note that the isomorphic relationship between $a$ and $b$
is first broken and then reestablished.) This connection poses a challenge for automation.

}

\vspace{2mm}
\noindent
{\bf On buggy examples.} We have deliberately injected a number of
different bugs into originally safe programs.  To name a few: wrongly
specified ``enclosure'' heap ({\sf *buggy-isocopy}), buggy recursive
definitions ({\sf *buggy-mark2}), buggy stack manipulating statements
({\sf *buggy-length}), and buggy heap-manipulating statements ({\sf
*buggy-mark1}).  For these cases, the performance of our verifier can
diverge significantly.  For most examples, we fail and terminate
quickly. Notably, however, for the case of {\sf *buggy-mark1}, our
entailment check procedure exhausts its options without being able to
find a successful proof.



\balance

\section{Further Related Work and Discussion}
\label{sec:related}

It is possible, but very difficult, to reason in Hoare logic about
programs with pointers;
\cite{morris82tfpm,bornat00mpc} explore this direction.
The resulting proofs are inelegant and remain too low-level to
be widely applicable, let alone being automated.

Separation Logic (SL) \cite{ohearn01local,reynolds02separation} was a
significant advance with local reasoning via a frame rule, influencing
modern verification tools.  For
example, \cite{berdine05smallfoot,botincan9smt,jacobs11verifast}
implement SL-based symbolic execution, as described
in~\cite{berdine05symexec}.  But there was a problem in accommodating
data structures with \emph{sharing}.

Bornat et al. \cite{bornat04space} present a pioneering SL-based
approach for reasoning about data structures with intrinsic
sharing. The attempt results in ``dauntingly
subtle'' \cite{bornat04space} definitions and verifications.  Thus it
is unclear how to automate such proofs.

Explicit naming of heaps naturally emerged as extensions of
SL \cite{duck13heaps}.  Reynolds \cite{reynolds03course} conjectured
that referring explicitly to the current heap in specifications would
allow better handles on data structures with sharing.
We advance  \cite{duck13heaps} with a proof method for 
propagating and reasoning about recursive
definitions.  More specifically, we now considered \emph{entailments}
between such definitions, whereas \cite{duck13heaps} only considered
simple safety properties, which can be translated to the
satisfiability problem restricted to non-recursive definitions. But
more importantly, it is this current paper that fully realizes
Reynolds' conjecture by connecting the explicit subheaps to the global
heap ($\cal{M}$) with the concept of heap reality and formalizing the
concepts of ``evolution'' and ``enclosure''.  This leads to a new
frame rule, and consequently enables local reasoning.

\floatstyle{plain}
\restylefloat{figure}

\begin{wrapfigure}{r}{0.2\textwidth}
\begin{center}
\vspace{-3mm}
\includegraphics[width=0.18\textwidth]{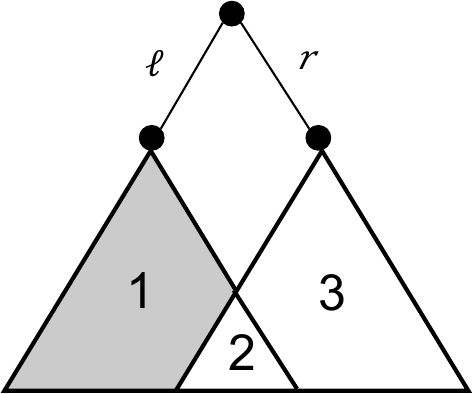}
\vspace{-3mm}
\caption{{\tt mark} DAG }
\label{fig:mark-dag}
\end{center}
\end{wrapfigure}

Next consider \cite{hobor13ramify} which addressed sharing (but not
automation).  Recall the {\tt mark} function, but now consider its
application on a DAG, Fig.~\ref{fig:mark-dag}.  The ``ramify'' rule
in \cite{hobor13ramify} would isolate the shaded heap portion {\tt 1}
and prove that the portion {\tt 1} has all been marked.  With the help
of the {\em magic wand}, this seems general.  Its application,
however, is counter-intuitive and hard to automate, because the
portion {\tt 1} is \emph{artificial}: it does not correspond to the
actual traversal of the code.


%



We now elaborate on our earlier comments about DF/IDF.
\ignore{
After SL, we have the method of \emph{dynamic frames}
\cite{kassios06dynamic} (DF), and later, the refinement to
\emph{implicit} dynamic frames (IDF) ~\cite{smans09implicit}.
Essentially, a DF is a mathematical expression describing a (super)set
of memory locations that a method refers to, a.k.a. the ``footprint''
of the method.  It is used in the specification phase of verification:
together with the provided pre and postcondition of a method, the DF
represents the ``modifies'' property of the method.  The big advantage
that DF/IDF brings over SL is both in expressiveness (frames can be
specified at a higher resolution) and automation (the VC's can be
generated in a particular way so as to be able to be automatically
dispensed, perhaps with some manual intervention).
}
In DF, the method footprint is described by a distinguished variable
in an expression containing program variables, and additional ghost variables.  
To have \emph{footprint compliance}, code is annotated with ghost variable statements.
An advantage of DF is in its expressivity of the footprint.
%
%
%
%
\ignore{
Consequently, DF systems
need to deal with \emph{footprint compliance}:
how to navigate the code and then produce appropriate
verification conditions (VC's) which guarantee that the footprint
encloses all memory accesses within the method.
For example, in some verifiers, e.g., Dafny
\cite{leino10dafny2}, ghost variables are used to explicitly describe
the dynamic frame, and the code may be annotated with ghost variable
assignments.  Correctness then requires that the heap updates are
enclosed in the distinguished ghost variable nominated as the dynamic
frame of the code.  A disadvantage is the added verbosity required on
the ghost variable expressions, and the added risk of bugs in matching
these expressions against program variable expressions.
This entails that the code needs to be annotated in order
to describe the behavior of the ghost variables 
(which do not appear in the original code).
ie: prone to bugs and verbosity,
to connect the ghost variables to the program variables.
}
IDF is a refinement of DF that avoids the need in DF to 
explicitly specify and then verify frame properties.
Instead, these properties are \emph{inferred} from assertions about
accessibility, and these are typically written in the pre and post
conditions of methods.
The concept of separation is then covered by reverting back to 
SL, ie. using separating conjunction.
As a result, IDF can concisely reason about frames.

Some prominent verifiers that use DF/IDF are Vericool \cite{smans08fase},
Verifast \cite{jacobs11verifast}, Dafny \cite{leino10dafny2}, Chalice
\cite{leino09chalice} and Viper \cite{viper}.
None of these systems directly support recursive data structures
in a general way.  As \cite{kassios11tutorial} remarks,
`` it is not clear how to prove properties involving footprints ...
, e.g., footprint compliance and self-framing, 
without an \emph{exact} non-recursive definition of dynamic frames. 
Such a definition however would in general require reachability predicates ...
Reachability predicates are not expressible in first order
logic, and therefore cannot be used with an SMT-based tool''.

\ignore{
Examples of works on explicit heaps are (Implicit) Dynamic Frames
[14,22] and Region Logic [1,2]. The underlying approach is to
represent the heap M as a (possibly implicit) total map over all
possible addresses, and to represent access or modification rights as
sets of addresses F. Separation is represented as set disjointness,
i.e., F1 ∩F2 = ∅. In [2], a new form of frame condition specifies
write, read, and allocation effects using region expressions; this
supports a “frame rule” that allows a command to read state on which
the framed predicate depends.

The approaches of DF and IDF require quite different answers to this,
but there is a commonality: the DF expression contains, in addition to
program variables, additional \emph{existentially quantified}
variables.  These may be (a) explicitly written as \emph{new}
variables not appearing in the program (eg. Dafny \cite{}) or (b)
implicitly due to the use of predicates with \emph{recursive
definitions}.  An example DF: a list rooted at (program variable) $x$
is either the empty set of memory locations, or \emph{there exists a
location} $next$ such that $x$ is a memory location for a pair $(val,
next)$ and DF comprises $x$ and the list rooted at $next$.  Note that
in case (b), there is an \emph{unbounded} number of existentially
quantified variables.  Hereafter we abbreviate existentially quantified
with ``ghost''.  The key to footprint compliance is then
to \emph{connect} the ghost variables to the program variables in the
method that actually traverse the data structure.  The general
approach has to \emph{annotate} the program.
{\color{magenta}
Examples ... Dafny: code on ghosts (bugs!), Viper (explicit fold/unfold)

Some prominent verifiers that use DF/IDF are Vericool \cite{smans08fase},
Verifast \cite{jacobs11verifast}, Dafny \cite{leino10dafny2}, Chalice
\cite{leino09chalice} and Viper \cite{viper}.
}
}

The work \cite{nanevski09structuring} shows that by choosing less
straightforward definitions of heaps and of heap union in {\sf Coq},
we can obtain effective reasoning with abstract heap variables, and
hence support full separation logic without resulting in excessive
proof obligations.  As a result, proofs of a number of simple but
realistic programs have been successfully mechanized.  Similarly, the
work \cite{sergey15pldi}, which described a mechanized proof of a
concurrent in-place spanning tree construction algorithm, bears
resemblance to our graph marking example.  This is because they
traverse via two recursive calls (but they are unconcerned about their
relative order).  Therefore this work does address the challenge of
dealing with the interaction of two recursive calls.  Both these
works \cite{nanevski09structuring,sergey15pldi} do not address the
automation of local reasoning.

Next we briefly
mention some recent work on Region Logic, see
e.g. \cite{banerjee08regionallogic}.  This work is related to DF: it
is essentially a form of Hoare logic for object-based programs.  A
region, like a dynamic frame, is an expression to describe the
footprint of a function. Finally, we also mention \cite{lee11cav,li15sas} 
that address sharing in the context of shape analysis.
In contrast, our current paper focuses on functional verification, thus 
the ability to perform strongest postcondition propagation is crucial. 
In this context, automated framing of such assertion formula is 
a non-trivial task.

\vspace{2mm}
\noindent
{\bf On ``Proof by Framing'':} This paper advances local reasoning
when dealing with data structures with sharing. However, it should be
noted that local reasoning might not always be applicable. Consider the
following example: a modification of the {\tt mark} example, but
instead working on DAGs.

\vspace{2mm}
\begin{center}
\stuff{
void countpath(struct node $*$x) \{ \\
 \>     if (!x) return; \\
 \>     struct node $*$l = x->left, $*$r = x->right; \\
 \>     x->mark = x->mark + 1;  \\
 \>     countpath(l); countpath(r); \\
 \}
}
\end{center}
\vspace{2mm}

\noindent
This program, intuitively, counts the number of ``paths'' from the
root to each node in the DAG. It cannot be verified using our frame
rule(s), simply because the sets of cells modified by left and right
recursive calls overlap: what established by the first cannot be
framed over the subsequent fragment. In this case, it is
clear that ``local reasoning'' with framing is \emph{not} the way to
proceed. (It does not mean that we cannot prove such program using a
manual or a non-compositional method.)

\ignore{
Instead, we believe that to still achieve compositional reasoning,
``summary reuse'' is a potential solution, where the ability
to state the summary of heap transformations using
explicit heaps becomes highly advantageous over
traditional SL. Pursuing this direction is left as future work.
}


\section{Conclusion}
\label{sec:conclusion}
We considered the problem of automatically verifying programs which
manipulate complex data structures.  These structures, which may
exhibit unrestricted sharing including being cyclic, are defined in
our specification language which uses recursive definitions.  A key
feature of our definitions is their use of explicit heaps in
order to frame away constituent substructures, in preparation for local
reasoning.  


Our main contribution is then, given a program which has been annotated with
preconditions and postconditions, a method to:

\begin{itemize}
\item generate verification conditions via symbolic execution (which realizes 
strongest postcondition reasoning) and framing; and 
\item discharge the verification conditions by using a standard method of
unfolding recursive definitions.
\end{itemize}

\noindent
Finally, we presented a prototype implementation and
demonstrated it over a number of representative programs.  In
particular, we focused on a graph marking program and presented its
first verification by systematic means.  An practical outcome of this
important example is that its proof provides a template for the formal
proof of a class of programs which traverse possibly cyclic data structures.


\clearpage

\bibliographystyle{plain}
\bibliography{references}



\end{document}